\documentclass[12pt, draftclsnofoot, onecolumn]{IEEEtran}
\usepackage[compress]{cite}
\usepackage{graphicx}
\usepackage{color}
\usepackage{amsmath}
\usepackage{hyperref}
\usepackage{amssymb}
\usepackage{amsfonts}
\usepackage{psfrag}
\usepackage{booktabs}
\usepackage{array}
\usepackage{subfig}
\usepackage{caption}
\usepackage[named]{algo}
\usepackage{makecell}
\usepackage{algorithmic}
\usepackage{algorithm}
\usepackage{algorithm,setspace}
\usepackage{tikz}
\usepackage{pgfplots}
\usepackage[export]{adjustbox}
\usepackage{verbatimbox}
\usepackage{stackengine}

\IEEEoverridecommandlockouts

\newcommand{\etal}{\textit{et al.}}

\newcommand\raisepunct[1]{\,\mathpunct{\raisebox{0.4ex}{#1}}}

\usepackage[colorinlistoftodos]{todonotes}
\usepackage{acronym}
\acrodef{2D}{2-dimensional}
\acrodef{3D}{3-dimensional}
\acrodef{4G}{4th generation}
\acrodef{5G}{5th generation}
\acrodef{6G}{6th generation}
\acrodef{ADT}{angular diversity transmitter}
\acrodef{ADR}{angular diversity receiver}
\acrodef{ANSI}{American national standards institute }
\acrodef{AP}{access point}
\acrodef{AWGN}{additive white Gaussian noise}
\acrodef{AWGR}{high port count arrayed waveguide grating router}
\acrodef{BER}{bit error ratio}
\acrodef{CDF}{cumulative distribution function}
\acrodef{CSI}{channel state information}
\acrodef{CPC}{compound parabolic concentrator}
\acrodef{CoMP}{coordinated multipoint}
\acrodef{DC}{direct current}
\acrodef{DCO{-}OFDM}{direct current-biased optical orthogonal frequency division multiplexing}
\acrodef{DFB}{distributed feedback laser}
\acrodef{EGC}{equal gain combining}
\acrodef{FEC}{forward error correction}
\acrodef{FFT}{fast Fourier transform}
\acrodef{FF}{fill factor}
\acrodef{FOV}{field of view}
\acrodef{FSO}{free space optics}
\acrodef{IEC}{International electrotechnical commission}
\acrodef{IFFT}{inverse fast Fourier transform}
\acrodef{IM{-}DD}{intensity modulation and direct detection}
\acrodef{IR}{infrared}
\acrodef{ISI}{inter-spot interference}
\acrodef{ICI}{inter-cluster interference}
\acrodef{JT}{joint transmission}
\acrodef{LD}{laser diode}
\acrodef{LED}{light-emitting diode}
\acrodef{LiFi}{light fidelity}
\acrodef{MPE}{maximum permissible exposure}
\acrodef{MHP}{most hazardous position}
\acrodef{MRC}{maximum ratio combining}
\acrodef{MUI}{multi-user interference}
\acrodef{NIR}{near infrared}
\acrodef{NOMA}{non-orthogonal multiple access}
\acrodef{PAM}{pulse amplitude modulation}
\acrodef{PIN}{positive-intrinsic-negative}
\acrodef{PSD}{power spectral density}
\acrodef{PD}{photodiode}
\acrodef{OFDM}{orthogonal frequency division multiplexing}
\acrodef{OFDMA}{orthogonal frequency division multiple access}
\acrodef{OOK}{on-off keying}
\acrodef{OW}{optical wireless}
\acrodef{OWC}{optical wireless communication}
\acrodef{QAM}{quadrature amplitude modulation}
\acrodef{RF}{radio frequency}
\acrodef{RIN}{relative intensity noise}
\acrodef{RSS}{received signal strength}
\acrodef{SC}{selection combining}
\acrodef{SWC}{switched combining}
\acrodef{SDMA}{space division multiple access}
\acrodef{SIC}{successive interference cancellation}
\acrodef{SINR}{signal-to-interference-plus-noise ratio}
\acrodef{SNR}{signal-to-noise ratio}
\acrodef{TIA}{transimpedance amplifier }
\acrodef{TEM}{transverse electromagnetic}
\acrodef{VCSEL}{vertical cavity surface emitting laser}
\acrodef{VLC}{visible light communication}
\acrodef{WiFi}{wireless fidelity}

\begin{document}
\title{High-Speed Imaging Receiver Design for\\ 6G Optical Wireless Communications: \\ A~Rate-FOV Trade-Off} 

\vspace{-0.3cm}
\author{\IEEEauthorblockN{\normalsize{Mohammad Dehghani Soltani, Hossein Kazemi, Elham Sarbazi, Taisir E. H. El-Gorashi, Jaafar M. H. Elmirghani, Richard V. Penty, Ian H. White, Harald Haas and Majid Safari}}
\vspace{-0.3cm}
\thanks{Part of this work is accepted to be presented at the 7th IEEE ICC Workshop on Optical Wireless Communications (OWC'22), 16–20 May 2022 \cite{MDS2022ICC}.}
}

\maketitle

\vspace{-1.5cm}

\begin{abstract}
\vspace{-0.2cm}
The design of a compact high-speed and wide \ac{FOV} receiver is challenging due to the presence of two well-known trade-offs. The first one is the area-bandwidth trade-off of photodetectors (PDs) and the second one is the gain-FOV trade-off due to the use of optics. The combined effects of these two trade-offs imply that the achievable data rate of an imaging optical receiver is limited by its FOV, i.e., a rate-FOV trade-off.
To control the area-bandwidth trade-off, an array of small PDs can be used instead of a single PD. Moreover, in practice, a large-area lens is required to ensure sufficient power collection, which in turn limits the receiver \ac{FOV} (i.e., gain-FOV trade-off). In this paper, we propose an imaging receiver design in the form of an array of (PD) arrays. To achieve a reasonable receiver \ac{FOV}, we use individual focusing lens for each PD array rather than a single collection lens for the whole receiver. The proposed array of arrays structure provides an effective method to control both gain-FOV trade-off (via an array of lenses) and area-bandwidth trade-off (via arrays of PDs). 
We first derive a tractable analytical model for the signal-to-noise ratio (SNR) of an array of PDs where the maximum ratio combining (MRC) has been employed. Then, we extend the model for the proposed array of arrays structure and the accuracy of the analytical model is verified based on several Optic Studio-based simulations.
Next, we formulate an optimization problem to maximize the achievable data rate of the imaging receiver subject to a minimum required \ac{FOV}. 
%
The optimization problem is solved for two commonly used modulation techniques, namely, on-off keying (OOK) and direct current (DC) biased optical orthogonal frequency division multiplexing (DCO-OFDM) with variable rate quadrature amplitude modulation (QAM). It is demonstrated that a data rate of $\sim 24$~Gbps with a FOV of $15^{\circ}$ is achievable using OOK with a total receiver size of $2$~cm $\!\times\!$ $2$~cm.
\end{abstract}

\vspace{-.4cm}

\begin{IEEEkeywords}
\vspace{-0.2cm}
Photodetector, PIN photodiode, lens, array and optical wireless communication.
\vspace{-0.3cm}
\end{IEEEkeywords}

\section{Introduction}
\vspace{-0.3cm}
\label{Section1}
The global demand for high-speed data rate is increasing rapidly, where no saturation trends have yet been observed \cite{AIR_2020}. Three-dimensional (3D) holographic displays with the ability to create virtual reality (VR) environments, $8$K ultra-high-definition (UHD) video streaming and online video games are just few examples of applications that require high-speed data rate.
It is forecasted there will be $5.7$ billion connected mobile users by 2023 \cite{CiscoInternet_2023}. This yields an exponential growth in data traffic of wireless communications.
The radio frequency (RF) may struggle to handle this high data traffic. In fact, RF spectrum is getting congested and wireless devices are seriously interfering with each other. As a result, the data throughput of the devices is affected severely, and the established connection may be poor.


Optical wireless communication (OWC), which permits the availability of a huge spectrum, is a propitious technology to alleviate the RF spectrum congestion by offering solutions such as free space optical (FSO) communications \cite{FSOKhalighi2014}, (ii) visible light communications (VLC) \cite{karunatilaka2015led}, (iii) underwater OWC \cite{UnderwaterZeng2017}, (iv) optical camera communications (OCC) \cite{OCCShivani2017} and (v) wireless networking with light, which is also referred to as LiFi\cite{LiFi2016,OWCHaas2020,Koonen2018JLT}. Moreover, OWC can create smaller communication cells in the network, known as attocell \cite{LiFi2016} which is a vital key to unlock the path to exponential capacity growth for indoor scenarios \cite{Elgala2011Mag}.
OWC can offload heavy traffic loads from congested RF wireless networks, therefore making the RF spectrum to be available for emerging wireless applications such as the Internet of Things (IoT). 


High-speed aggregate Terabit per second indoor OWC systems are being realized thanks to the large modulation bandwidth of laser diodes \cite{Hong:21:Tbps,OBrienTerabit2018,KoonenHighSpeed2020,Sarbazi2000PIMRC,Kazemi2000PIMRC}.
Hong \etal\ demonstrated that a ten-channel wavelength division multiplexing (WDM) intensity modulation/direct detection (IM/DD) system can achieve $>1$ Tb/s capacity at a perpendicular distance of $3.5$ m with a lateral coverage up to $1.8$ m \cite{Hong:21:Tbps}.
A $1$ Tbit/s bi-directional free-space connection has been presented in \cite{OBrienTerabit2018}, which employs ten WDM channels with pulse amplitude modulation (PAM)-$4$. The coverage area of $2.54$ m$^2$ is reported.
Beam-steering laser-based optical multiuser systems, which is capable of supporting up to $128$ beams carrying up to
$112$ Gbit/s per beam has been developed in \cite{KoonenHighSpeed2020}. 

One of the major challenges for the full adoption of OWC in 6G indoor wireless networks is to design high-speed optical receivers that can operate reliably in a mobile environment. Such receiver requires small detectors with large bandwidth and a large \ac{FOV}\footnote{It is noted that throughout this paper, \ac{FOV} is defined based on the full-cone angle.} to support mobility. However, enabling a high data rate and a large \ac{FOV} at the same time is challenging due to the following reasons. Firstly, there is a trade-off between photodetector (PD) bandwidth and its sensitive area. The utilization of light-focusing optics (lens or concentartor) can compensate for the small area of a high-bandwidth PD and enhance the received \ac{SNR}. However, the use of optics gives rise to  a second trade-off between gain and the \ac{FOV} of the receiver in accordance with etendue law of conservation \cite{chaves2008introduction}. This indicates that the improved \ac{SNR} is achieved in expense of a narrow \ac{FOV} receiver.

The choice of angle diversity receiver (ADR), which consist of multiple narrow-\ac{FOV} detectors facing different directions, is one way to achieve a larger aperture and wider \ac{FOV} \cite{Kahn97WirelessInfrared}.
In \cite{Alkhazragi:21}, a wide-\ac{FOV} receiver using fused fiber-optic tapers is proposed, which comprises hundreds of thousands of tapered optical fibers. An overall \ac{FOV} of $30^{\circ}$ and optical gain of $~121.3$ are reported. However, receiver designed based on these approaches are not as compact as typical imaging receivers. 
Fluorescent optical concentrators are also introduced to enhance the receiver \ac{FOV}, which are mainly applicable for visible light communications\cite{CollinsFOV2014,Riaz2019Fluorescent,Manousiadis:WideFOV16}.
A fluorescent concentrator is shown to achieve a gain of $50$ times higher than the conventional concentrators with the same \ac{FOV} \cite{CollinsFOV2014}. In \cite{Manousiadis:WideFOV16}, a threefold increment in data rate is reported by using a fluorescent antenna that can support \ac{FOV} of $120^{\circ}$. However, the maximum data rates reported using this approach are still limited since color conversion restricts the bandwidth.

An array of PDs along with proper optics is another promising solution which can offer an enhanced \ac{SNR} and a wide \ac{FOV} while ensuring a compact design, which is essential for portable mobile devices. 
Detector arrays can be designed using PIN or avalanche photodiodes (APD), which typically operate in thermal noise limited and shot noise limited regimes, respectively \cite{StephenBAlexander1997}. For mobile scenarios, PIN photodetectors are more preferable due to low biasing voltage \cite{KHARRAZ20131493}. The output current of \ac{PIN} photodetectors are typically so small that this necessitates the use of an amplifier. Among various types of amplifiers, \acp{TIA} are best suited to most optical wireless applications, since they support a large dynamic range and have a wide bandwidth without the need for an equalizer \cite{personick1973receiver, Kahn97WirelessInfrared}. 
High speed TIAs have been designed and fabricated in many studies \cite{Bashiri26GHzTIA2010, Kim40GHzTIA2019, Gracia60GHzTIA2017, Garcia66GHzTIA2018, Vasilakopoulos96GHzTIA2015}. For instance, TIAs with $66$ GHz and $96$ GHz bandwidth are demonstrated in \cite{Garcia66GHzTIA2018} and \cite{Vasilakopoulos96GHzTIA2015}, respectively.
The other consideration when designing an detector array receiver is the choice of a combining technique, e.g., equal gain combining (EGC), selection combining (SC), switched combining (SWC) or the maximal ratio combining (MRC).  
It has been mathematically proven that even though an array of PDs augments more noise, employing an efficient combining algorithm  can help us achieve a superior probability of error performance compared to a single-detector receiver in practical channel conditions \cite{Bashir2020FSO,Tsai2021single}.

Several experimental works in the literature have demonstrated the use of an array of PDs to achieve high data rates. 
In \cite{Koonen:2020:Receiver}, a broadband receiver with a large aperture is introduced. The authors experimentally demonstrated a data rate of $1$ Gbps at a \ac{FOV} of $10^{\circ}$. Umezawa \etal\ have reported an array of $8\times 8$ pixels along with a $15$ mm diameter-lens, which manifests a \ac{FOV} of $6^{\circ}$ in \cite{Umezawa2022LargeSubmil}. They have achieved data rate of $25$ Gbps at a distance of $10$ m.
Signal detection enhancement for PD arrays has been discussed in \cite{Vilnrotter2002Adaptive}. 
Vilnrotter \etal\ have proposed an optimum array detection algorithm, which requires intensive computation and a simpler suboptimum structure that can achieve performance improvement up to $5$ dB. All these studies emphasize the realization of high-speed mobile receiver by means of an array of PDs. However, to the best of our knowledge, there is no tractable analytical framework in the literature enabling imaging receiver designs that can optimally explore the trade-off between achievable data rate and FOV. In this study, we initially derive a general \ac{SNR} model for a single array, where MRC technique is employed. Then, we extend the model for an array of arrays. Afterwards, we develop an optimization problem to design an imaging receiver with maximum achievable data rate supporting a required \ac{FOV}. In summary, the contributions of this paper are pointed out as follows:
\begin{itemize}
\item[$\bullet$] In-depth analysis of an imaging optical receiver by modeling beam spot size considering both focused and defocused designs for the imaging receiver.
\item[$\bullet$] Development of a tractable framework for \ac{SNR} of a static single array as well as the average \ac{SNR} over different receiver tilt angles. 
\item[$\bullet$] Formulating an analytical optimization problem to achieve an optimum design for a receiver that can support required constraints such as a minimum \ac{FOV} and maximum bit error rate (BER). The optimization problem has been evaluated for \ac{DC} biased optical orthogonal frequency division multiplexing (DCO-OFDM) with variable rate quadrature amplitude modulation (QAM) and on-off keying (OOK).
\item[$\bullet$] Validating analytical models and derivations using realistic simulations conducted in OpticStudio (Zemax) software.  
\item[$\bullet$] Presenting insightful results and detailed discussions about the impact of various trade-offs on receiver design, in the presence of non-idealities and practical aspects of the imaging receiver components.
\end{itemize}


The rest of this paper is organized as follows. In Section~\ref{ReceiverArchitecture}, the area-bandwidth and gain-\ac{FOV} trade-offs are discussed. In Section~\ref{SNRAnalysis}, we present analytical framework of \ac{SNR} for a single array as well as an array of arrays. In Section~\ref{Sec: Optimum Geometric Design}, the optimization problem is formulated to obtain the optimum design of a high-speed receiver.  Simulation results are presented in Section~\ref{Sec:SimulationResults}. In Section~\ref{Sec:Summary}, summary and concluding remarks of this study are drawn and a number of possible directions are suggested for the future research.

\begin{figure}[t!]
 \centering
\includegraphics[width=0.32\textwidth]{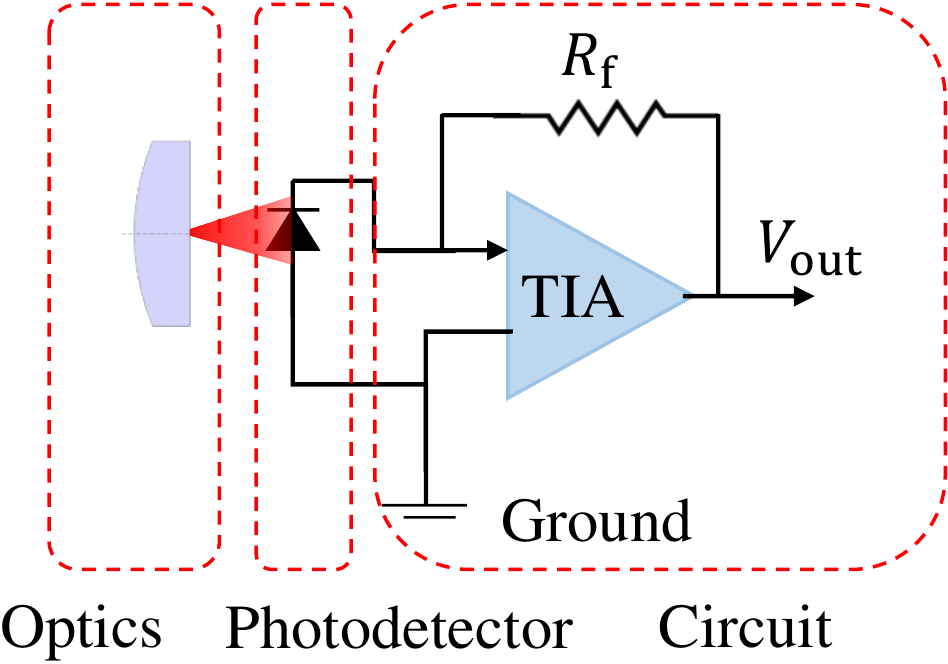} 
\caption{A common structure of an optical wireless receiver.}
\label{Fig-ReceiverStru}
\vspace{-0.8cm}
\end{figure}

\vspace{-0.3cm}
\section{Design Trade-offs for High-Speed OWC Receivers}
\vspace{-0.1cm}
\label{ReceiverArchitecture}
A common architecture of a receiver for OWC is shown in Fig.~\ref{Fig-ReceiverStru}, which includes three main elements: a \ac{PD}, optics and a \ac{TIA}. The proper selection of these elements plays a crucial role in addressing the design trade-offs as well as fulfilling the system requirements such as a desired data rate, \ac{FOV}, and \ac{BER}.  
High bandwidth PDs are more favorable for high-speed applications, however, the higher the bandwidth, the smaller the sensitive area. This is recognized as the area-bandwidth trade-off and states that a high bandwidth \ac{PD} may not be able to collect sufficient power, particularly when the beam spot radius at the receiver is much larger than the PD side length. Furthermore, choosing high bandwidth PDs in a power-limited regime is not a good practice as it adds more noise to the system, resulting in \ac{SNR} degradation. 
The use of imaging optics in front of the PD can help to collect more power and improve the \ac{SNR}. However, the \ac{FOV} of the receiver will be limited due to the law of conservation of etendue \cite{chaves2008introduction}, resulting in a gain-\ac{FOV} trade-off. 
Next, we will discuss the area-bandwidth and gain-\ac{FOV} trade-offs in more details. 


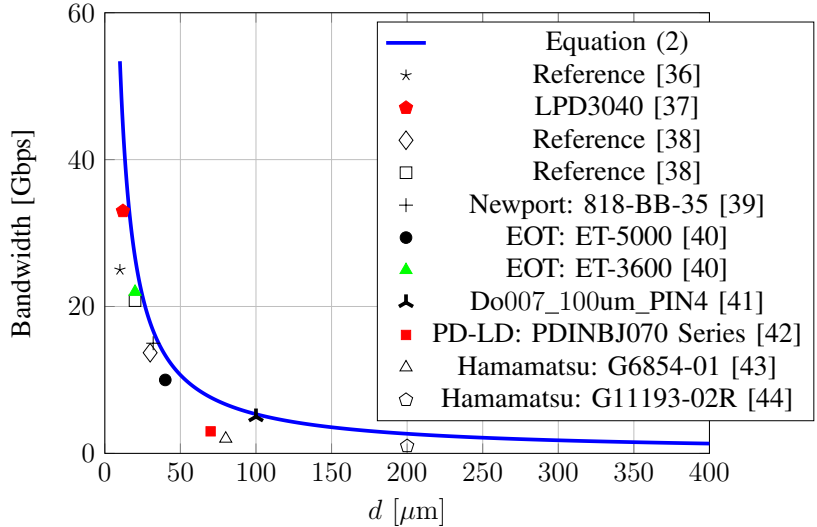
\begin{figure}[t!]
\begin{center}
\resizebox{.67\textwidth}{!}{
\begin{tikzpicture}
\begin{axis}[
        xlabel=$d\ {[\mu \rm{m}]}$,
        ylabel=Bandwidth {[Gbps]},
        xmin = 0, xmax = 400,
        ymin = 0, ymax = 60,
        grid = both,
        width = 0.65\textwidth,
        height = 0.5\textwidth,
        xtick distance = 50,
        ytick distance = 20,
        legend style={at={(0.45,0.98)},anchor=north west},
        ]
        \addplot[
               smooth, ultra thick,
               blue,
               ] file {AllData/BCt.dat};
        \addplot[black, only marks, mark = star, mark size = 2.5pt] plot coordinates {(10,25)};
        \addplot[red, only marks, mark = pentagon*, mark size = 3pt] plot coordinates {(12,33)};
       \addplot[black, only marks, mark = diamond, mark size = 4pt] plot coordinates {(30,13.7)};
       \addplot[black, only marks, mark = square, mark size = 2.5pt] plot coordinates {(20,20.8)};
       \addplot[black, only marks, mark = +, mark size = 3pt] plot coordinates {(32,15)};
       \addplot[black, only marks, mark = *, mark size = 2.5pt] plot coordinates {(40,10)};
       \addplot[green, only marks, mark = triangle*, mark size = 3pt] plot coordinates {(20,22)};
       \addplot[black, only marks, ultra thick, mark = Mercedes star, mark size = 3.5pt] plot coordinates {(100,5)};
       \addplot[red, only marks, mark = square*, mark size = 2pt] plot coordinates {(70,3)};
       \addplot[black, only marks, mark = triangle, mark size = 3pt] plot coordinates {(80,2)};
       \addplot[black, only marks, mark = pentagon, mark size = 3pt] plot coordinates {(200,1)};
        \legend{
        Equation \eqref{Bandwidth-SimplifiedEq},
        Reference \cite{Chen2019HighSpeed},
        LPD3040 \cite{DataSheetLPD3040},
        Reference \cite{Umezawa2018High},
        Reference \cite{Umezawa2018High},
        Newport: 818-BB-35 \cite{DataSheet15GHzNewport},
        EOT: ET-5000 \cite{DataSheetEOT},
        EOT: ET-3600 \cite{DataSheetEOT},
        Do$007\_100$um$\_$PIN4 \cite{DataSheetGCS},
        PD-LD: PDINBJ070 Series \cite{DataSheetPDLD},
        Hamamatsu: G6854-01 \cite{DataSheetHamamatsuG6854-01},
        Hamamatsu: G11193-02R \cite{DataSheetHamamatsuG11193-02R},
        }
        \end{axis}
\end{tikzpicture}
}
\end{center}
\vspace{-0.5cm}
\caption{Bandwidth versus the side length of a PIN photodetector, presenting the area-bandwidth trade-off. Several practical and commercial PIN PDs have been included. } 
\label{Fig-Bandwidth-PIN}
\vspace{-0.8cm}
\end{figure}

\vspace{-0.3cm}
\subsection{Area-Bandwidth Trade-off}
\vspace{-0.1cm}
\label{Sec:Area-Bandwidth Trade-off}
The bandwidth of PIN photodetectors is expressed as \cite{StephenBAlexander1997}:\vspace{-0.2cm}
\begin{equation}
\label{Bandwidth-General}
    B=\dfrac{1}{\sqrt{\left(2\pi R_{\rm s}\epsilon_0\epsilon_{\rm r}\dfrac{A}{\ell} \right)^2+\left( \dfrac{\ell}{0.44v_{\rm s}}\right)^2}},\vspace{-0.1cm}
\end{equation}
where $R_{\rm s}$ is the junction series resistance, $\epsilon_0$ is the permittivity in vacuum ($8.85\times 10^{-12}$ F.m$^{-1}$) and $\epsilon_{\rm r}$ is the relative permittivity of the semiconductor. The area and the length of the depletion region are denoted by $A$ and $\ell$, respectively. Also, $v_{\rm s}$ is the carrier saturation velocity which is limited by the hole saturation velocity. The junction series resistance, $R_{\rm s}$, typically varies from $7~\Omega$ to $9~\Omega$ depending on the \ac{PD} size \cite{Dupuis2014Exploring}. If the \ac{PD} is followed by a \ac{TIA} with an input resistance of $R_{\rm L}$, that is typically $50~\Omega$ for commercial products, the bandwidth is affected. In this case, $R_{\rm s}$ in \eqref{Bandwidth-General} should be replaced by the overall resistance $R_{\rm s}+R_{\rm L}$ \cite{StephenBAlexander1997}.


The left and right terms under the radical sign in \eqref{Bandwidth-General} represent the junction capacitance and transit time of a PIN detector, respectively. Clearly, as $\ell$ increases junction capacitance decreases, yielding a bandwidth increment, while the transit time increases, causing a bandwidth decrements. This trade-off results in an optimum length of depletion region which is denoted by $\ell_{\rm opt}$ \cite{StephenBAlexander1997}. 
Assuming a square shape for the PD with a side length of $d$, after some manipulations of \eqref{Bandwidth-General}, the optimal bandwidth at $\ell=\ell_{\rm opt}$ can be rewritten as:\vspace{-0.1cm}
\begin{equation}
\label{Bandwidth-SimplifiedEq}
    B=\dfrac{1}{C_{\rm t}d}, \ \ \ \ \ \ {\rm where}\ \ \ \ C_{\rm t}=\sqrt{\dfrac{4\pi R_{\rm s}\epsilon_0\epsilon_{\rm r}}{0.44v_{\rm s}}},\vspace{-0.1cm}
\end{equation}
This equation is an upper bound for bandwidth and other depletion regions where $\ell\neq \ell_{\rm opt}$ result in a lower value for bandwidth.

Fig.~\ref{Fig-Bandwidth-PIN} illustrates the bandwidth versus side length of a photodetector for $\ell=\ell_{\rm opt}$, which provides a theoretical upper bound on the performance of practical detectors. For comparison, we have included several commercial PDs in Fig.~\ref{Fig-Bandwidth-PIN}, which show the tightness of the theoretical bound.
Note that the area-bandwidth trade-off can be observed from these results, where as $d$ (or equivalently $A$) increases, the bandwidth decreases. When the received optical power is sufficiently high, the selection of high bandwidth PDs are preferred to achieve high data rates.  

\vspace{-0.3cm}
\subsection{Gain-FOV Trade-off}
\label{ReceiverOptics}
\begin{figure}[t!]
		\centering  
	\subfloat[\label{sub1:Fig-AsphricLensThorlab}]{%
		\resizebox{.2\linewidth}{!}{\includegraphics{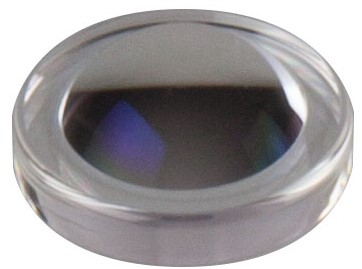}}
	} \quad\ \ 
	\subfloat[\label{sub2:Zemax}]{%
		\resizebox{.41\linewidth}{!}{\includegraphics{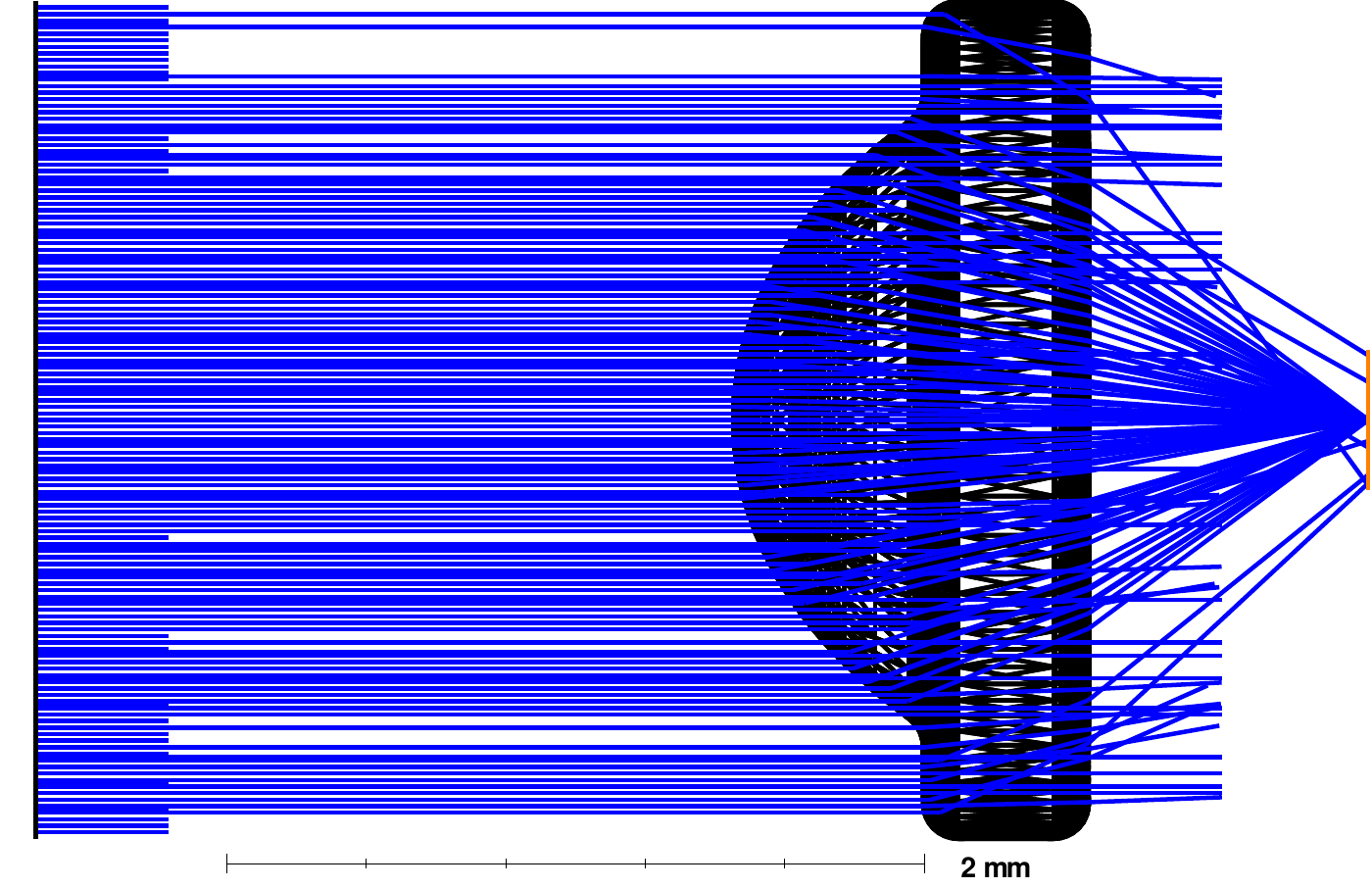}}
	}\\
	\vspace{-0.1cm}
	\caption{Thorlabs $354140$-B aspheric lens: (a) unmounted, (b) ray tracing simulations in OpticStudio \cite{Zemax}. }
	\label{Fig-LensSystem}
	\vspace{-0.5cm}
\end{figure}

The use of optics boosts the collected optical power at the expense of a reduced receiver \ac{FOV}, i.e., the gain-\ac{FOV} trade-off. 
In this paper, we focus on using an imaging lens and the study of \ac{CPC} is kept for future investigations. 
Without loss of generality, in this study, we employ an aspheric lens. We calculate the beam spot radius after the lens and the \ac{FOV} as a function of distance between the lens and the PD. These two functions are essential in the \ac{SNR} analysis and optimal design of receiver. 
Aspheric lenses are known to produce a precise and small beam spots. As an example, we consider a Thorlabs aspheric lens with the specifications given in Table~\ref{ThorlabAsphericTable} and the geometry shown in Fig.~\ref{Fig-LensSystem}. We note that the following theoretical analyses can be extended to other imaging lenses. 

\begin{table}[]
\centering
\caption[blah]{Thorlabs $354140$-B aspheric lens specifications \cite{DataSheetAsphericLens}.}
\label{ThorlabAsphericTable}
\begin{tabular}{|c|c|c|c|c|c|}
\hline
\begin{tabular}[c]{@{}c@{}}Effective \\ focal length\end{tabular} & Outer diameter & \begin{tabular}[c]{@{}c@{}}Back \\ focal length\end{tabular} & Clear aperture & Glass & \begin{tabular}[c]{@{}c@{}}refractive index \\ @ 850 nm\end{tabular} \\ \hline \hline
1.45 mm                                                           & 2.4 mm         & 0.82 mm                                                      & 1.6 mm         & D-ZK3 & 1.5809                                                               \\ \hline
\end{tabular}
\vspace{-0.6cm}
\end{table}

\begin{figure}[t!]
 \centering  
	\subfloat[\label{sub1:Fig-SpotSize}]{%
		\resizebox{.44\linewidth}{!}{\includegraphics{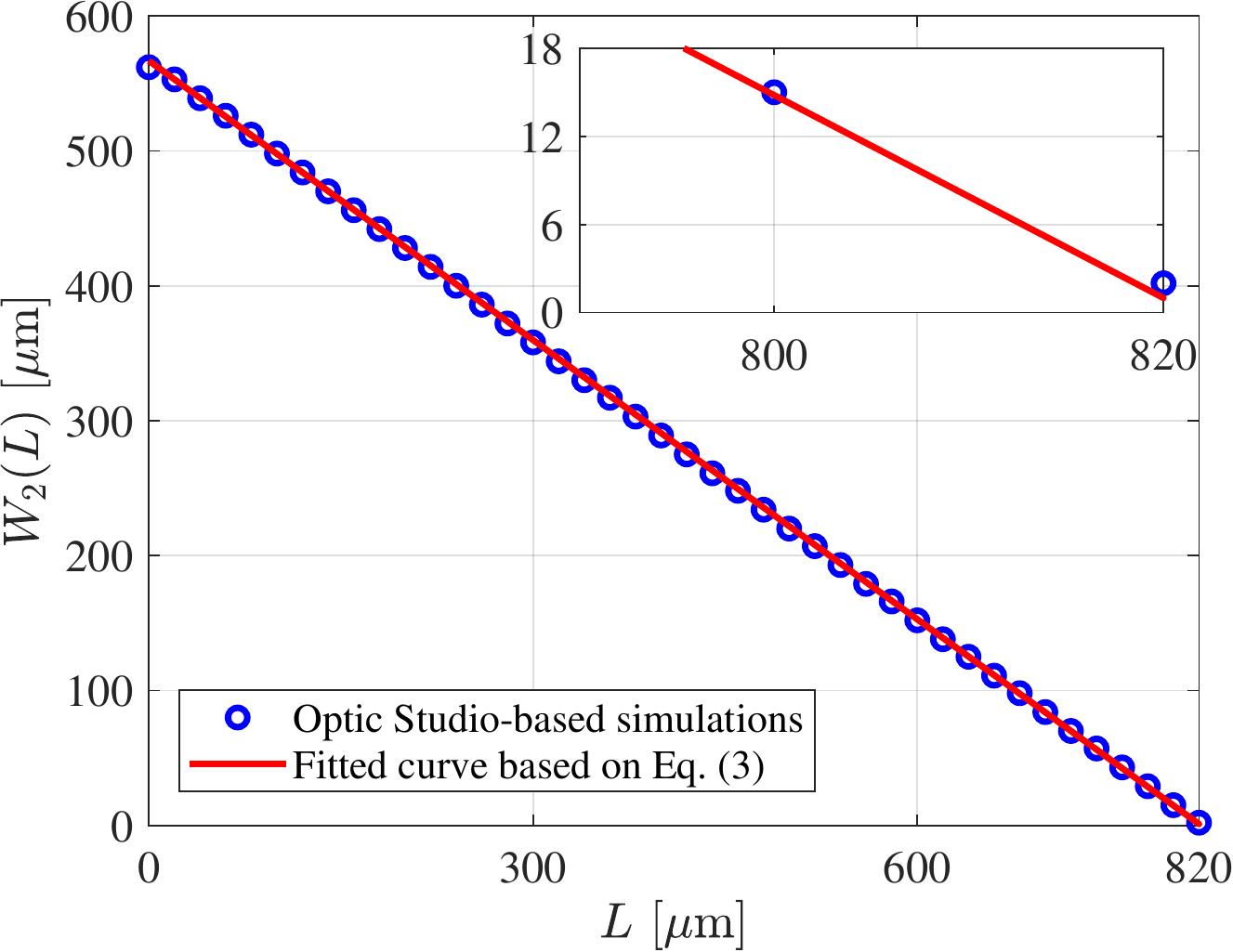}}
	} \quad\ \ 
	\subfloat[\label{sub2:Fig-SpotSize}]{%
		\resizebox{.45\linewidth}{!}{\includegraphics{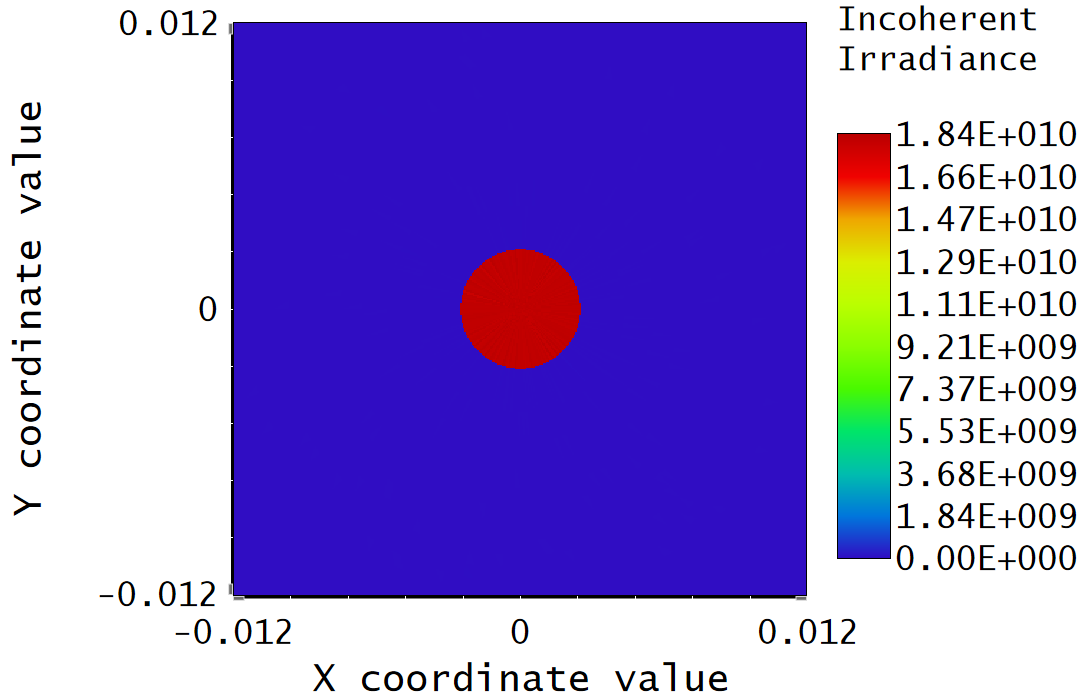}}
	}\\
	\vspace{-0.1cm}
\caption{(a) Beam spot radius versus distance from the lens obtained from OpticStudio software and the fitted curve with $b_1=0.69$ and $b_0=1\ \mu$m, (b) beam spot at $L=820\ \mu$m in OpticStudio (dimensions are in mm).}
\label{Fig-SpotSize}
\vspace{-0.8cm}
\end{figure}

Let $L$ denote the distance between the aspheric lens and the observing plane and $W_2(L)$ denote the beam spot radius after the lens. The fraction of the total beam power contained by $W_2(L)$ is denoted by $\eta$ in this study. Fig.~\ref{Fig-SpotSize} illustrates $W_2(L)$ at different values of $L$ for $\eta = 0.5$ the Thorlabs $354140$-B aspheric lens. The values of $W_2(L)$ are obtained using the commercial optical design software, OpticStudio \cite{Zemax}. 
According to Fig.~\ref*{sub1:Fig-SpotSize}, the beam spot radius can be well approximated by:
\vspace{-0.2cm}
\begin{equation}
\label{Eq:beamspotAftrLens}
    W_2(L)=b_1(f_{\rm b}-L)+b_0,\vspace{-0.2cm}
\end{equation}
where $f_{\rm b}$ is the back focal length of the aspheric lens. The two constants $b_0$ and $b_1$ can be obtained from the clear aperture of the lens denoted by $\rm{CA}$ and the spot radius in the diffraction limit. Diffraction limit determines the maximum resolution that can be achieved by an optical component. It is defined as $r_{\mathrm{DL}}=1.22 \lambda f_{\rm e}/\rm{CA}$, where  $\lambda$ is the wavelength of the light, and $f_{\rm e}$ is the effective focal length of the lens \cite{PrinciplesBorn1999}. 
Assuming that the incident beam spot is larger than the lens clear aperture, $W_2(L) = 0.5\rm{CA}$ at $L = 0$. Also, as shown in Fig.~\ref*{sub1:Fig-SpotSize}, at $L=820~\mu$m, $W_2(L)$ is equal to the diffraction limit. Therefore, $b_0$ and $b_1$ are obtained as:
\vspace{-0.3cm}
\begin{subequations}
\label{constants:BeamSpot}
\begin{align}
    &b_0=r_{\mathrm{DL}}\kappa,\\
    &b_1=\frac{\kappa}{2f_{\rm b}}\left({\rm CA}-2r_{\mathrm{DL}}\right),\vspace{-0.2cm}
\end{align}
\end{subequations}
where $f_{\rm e}$ and $\rm CA$ are the effective focal length and the clear aperture of the lens, respectively, which are provided in the datasheet \cite{DataSheetAsphericLens}. In \eqref{constants:BeamSpot}, $\kappa$ is a constant parameter, which is $\kappa=\sqrt{\eta}$. 

The thickness of the considered aspheric lens is $1$ mm, which is comparable to its radius of curvature that is $0.84$ mm \cite{DataSheetAsphericLens}. Therefore, the lens is not a thin lens\footnote{A thin lens is a lens with a thickness that is negligible compared to the radii of curvature of the lens surfaces \cite{saleh2019fundamentals}.}. 
We therefore evaluate the relation between \ac{FOV} and $L$ using the OpticStudio simulator. 
We accentuate that the traditional tangent equation that describes the relation between the \ac{FOV} and the effective focal length of the lens \cite{EdmundOpticsFoV} cannot be directly applied for all $0\leq L \leq f_{\rm b}$. 
There are several important remarks to highlight before introducing the approximation of the \ac{FOV}. Firstly, $L$ and the back focal point, $f_{\rm b}$, are measured from the back surface of the lens, while the effective focal length, $f_{\rm e}$, is measured from the principal plane of the lens. Secondly, the traditional expression of \ac{FOV} is applicable for $L$ that ensure the condition of $W_2(L)\leq d/2$, i.e., the beam spot is inscribed inside the PD. Thirdly, the \ac{FOV} is defined based on the half power metric. We can now present the approximation relation of FOV and $L$ which is valid for $W_2(L)\leq d$ as:
\begin{equation}
    {\rm{FOV}}=2\tan^{-1}\left(\dfrac{d}{2(L+f_{\rm e}-f_{\rm b})} \right), \ \ \ \ \ \ \ \   \dfrac{b_0-b_1f_{\rm b}-d/2}{b_1}\leq L\leq f_{\rm b},
\end{equation}
where the lower bound of $L$ is obtained by solving $2W_2(L)=d$ and using \eqref{Eq:beamspotAftrLens}. 

\begin{figure}[t!]
\centering
\begin{minipage}[b]{.4\textwidth}
\centering
\subfloat
  []
  {\label{Fig:PrThiltXL400}\includegraphics[width=1\textwidth,height=1.7cm]{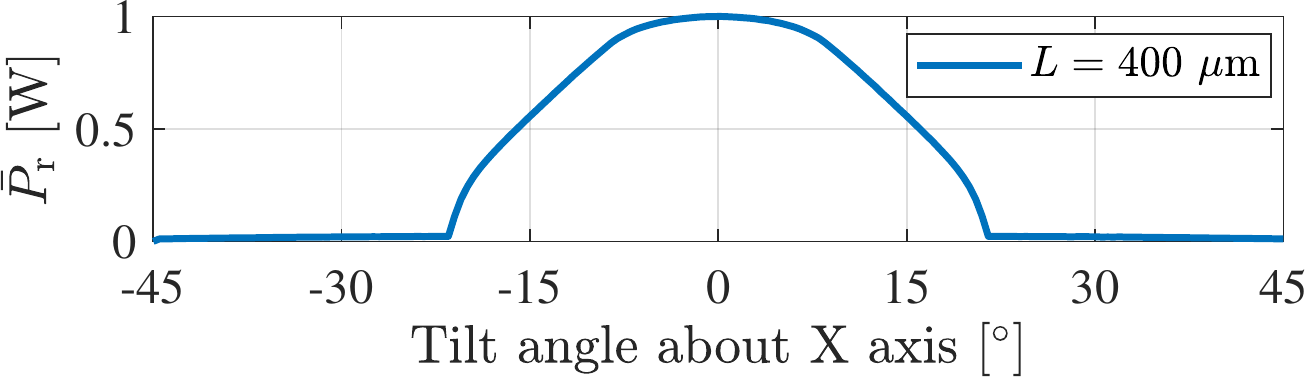}}
  
\subfloat
  []
  {\label{Fig:PrThiltXL600}\includegraphics[width=1\textwidth,height=1.7cm]{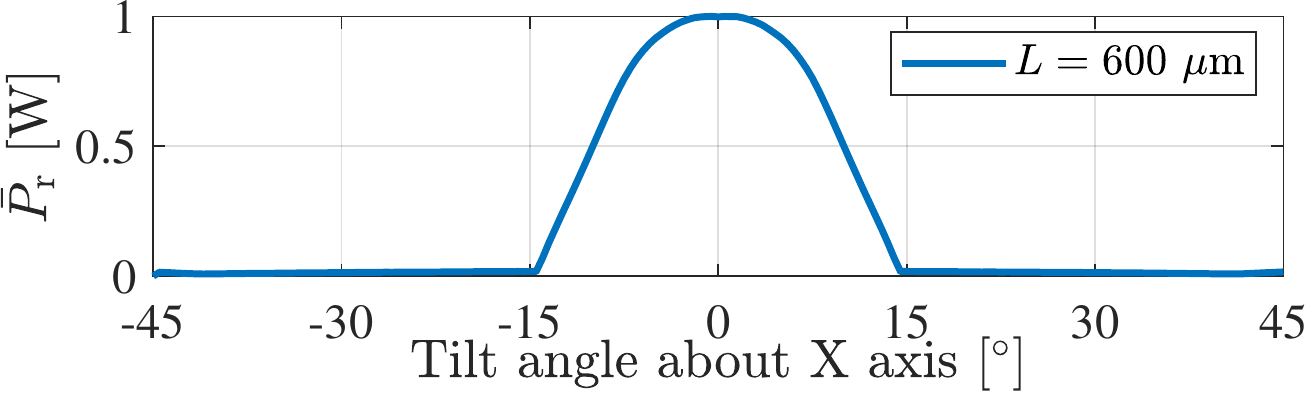}}

\subfloat
  []
  {\label{Fig:PrThiltXL820}\includegraphics[width=1\textwidth,height=1.7cm]{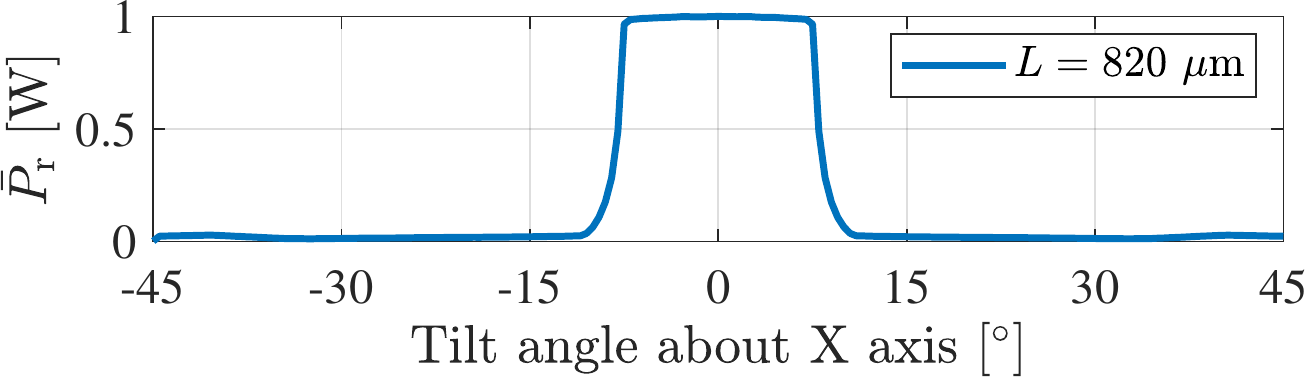}} 
\end{minipage}
\begin{minipage}[b]{.54\textwidth}
  \subfloat
    []
    {\label{Fig:L-FoV}\includegraphics[width=1\textwidth]{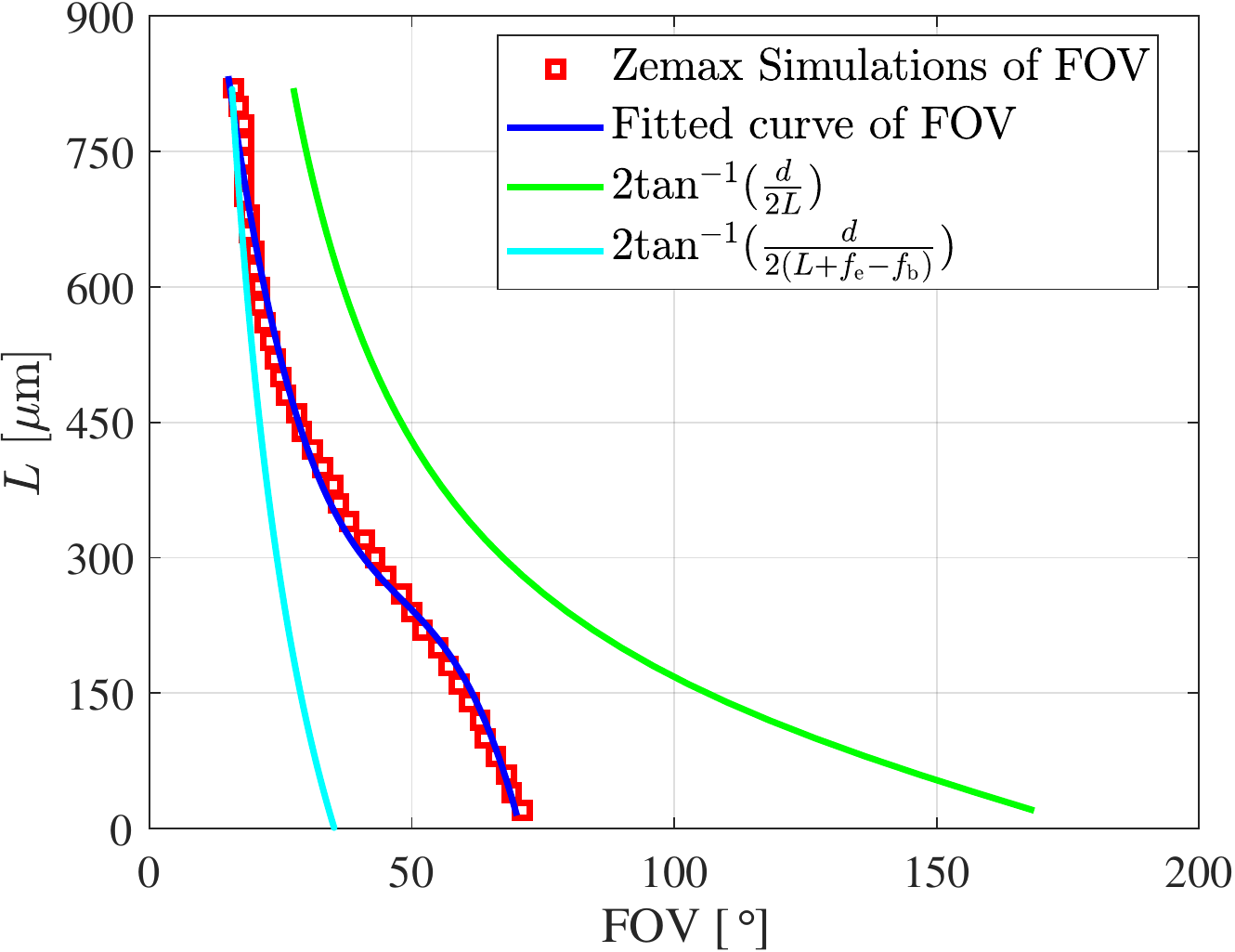}}
\end{minipage}
\vspace{-0.1cm}
\caption{Normalized received optical power versus transmitted beam tilt angle at (a) $L=400$ $\mu$m, (b) $L=600$ $\mu$m and (c) $L=820$ $\mu$m. (d) $L$ versus \ac{FOV} obtained from OpticStudio software, the fitted curve and the conventional analytical model for FOV. }
\label{fig:Test}
\vspace{-0.6cm}
\end{figure}

    
Fig.~\ref{Fig:PrThiltXL400}-\ref{Fig:PrThiltXL820} represent the normalized collected power at the PD, $\bar{P_{\rm r}}$, versus different tilt angles of transmitted beam about the $X$ axis. These results are obtained in OpticStudio and the collected power is normalized to the value at zero tilt angle.  
Three separate locations are selected to exhibit the effect of tilt angle on the collected power. It can be observed that at $L=400$ $\mu$m for the tilt angles between $-15.5^{\circ}$ to $15.5^{\circ}$ about the $X$ axis, the normalized power is greater than $0.5$. The range would be $[-10.5^{\circ},10.5^{\circ}]$ and $[-8^{\circ},8^{\circ}]$ at $L=600$ $\mu$m and $L=820$ $\mu$m, respectively. One typical definition of the \ac{FOV} is based on the half value of the collected power. Therefore, the \ac{FOV} at $L=400$ $\mu$m, $L=600$ $\mu$m and $L=820$ $\mu$m are $33^{\circ}$, $21^{\circ}$ and $16^{\circ}$, respectively. 
Fig.~\ref{Fig:L-FoV} illustrates the results of $L$ versus \ac{FOV} acquired in OpticStudio. A square PD with side length of $400$ $\mu$m is considered in these simulation results. The fitted curve to these results is:
\vspace{-0.2cm}
\begin{equation}
\label{Eq:L-FOV}
L=a_3 {\rm FOV}^3+a_2 {\rm FOV}^2+a_1 {\rm FOV}+a_0 ,\vspace{-0.2cm}
\end{equation}
where $(a_3,a_2,a_1,a_0)=(-0.08506,6.142,-159.5,1720)$ with a normalized root mean square error (NRMSE) of less than $10^{-4}$. The values for $(a_3,a_2,a_1,a_0)$ are obtained using the curve fitting toolbox of Matlab software. We have also plotted the approximation relation of FOV and $L$, which provides a good estimate for the range $520\ \mu\rm{m}\leq L\leq 820\ \mu\rm{m}$. For comparison, the curve of $2\tan^{-1}\left(d/2L\right)$ has been included, where we can see there is a noticeable gap between this curve and the OpticStudio simulations without applying the impact of $f_{\rm{e}}-f_{\rm{b}}$.


\vspace{-0.2cm}
\subsection{Array of Arrays Receiver Structure}
In an imaging optical receiver, the combined effect of the area-bandwidth and gain-FOV trade-offs introduces a new trade-off between the achievable data rate and FOV of the receiver, namely the rate-FOV trade-off. This is because for achieving higher data rate, more bandwidth (i.e., smaller PD) and higher collected power (i.e., larger collecting lens), which both imposing limits on the receiver FOV. In this paper, we propose an array of array structure for the imaging receiver that allows us to effectively control the area-bandwidth and gain-FOV trade-offs in order to achieve the desired system requirement, that is constrained by the rate-FOV trade-off.

 \begin{figure}[t!]
	\centering  
	\subfloat[\label{sub1:Receiver}]{%
		\resizebox{.49\linewidth}{!}{\includegraphics{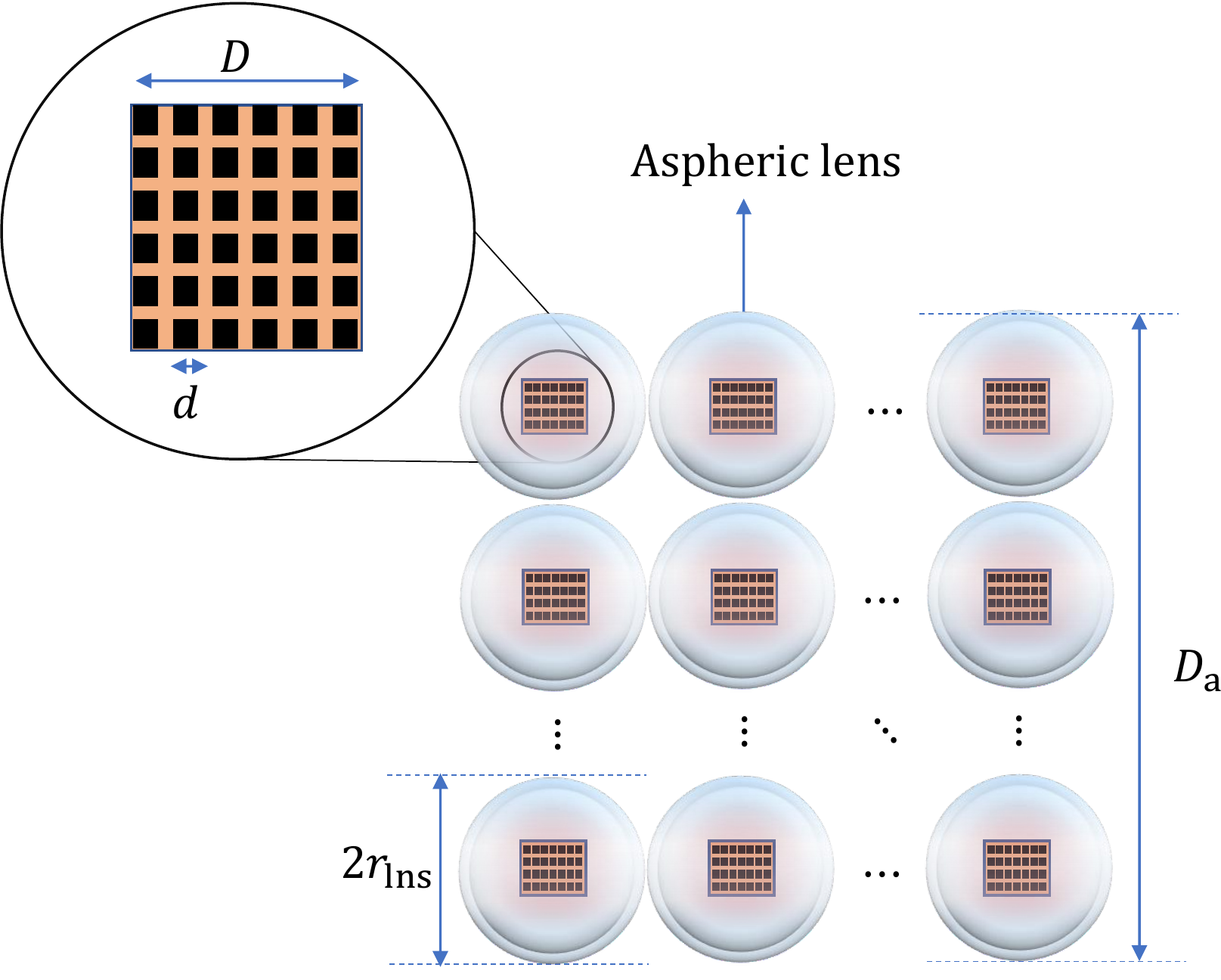}}}
	\subfloat[\label{sub2:Receiver}]{%
		\resizebox{.49\linewidth}{!}{\includegraphics{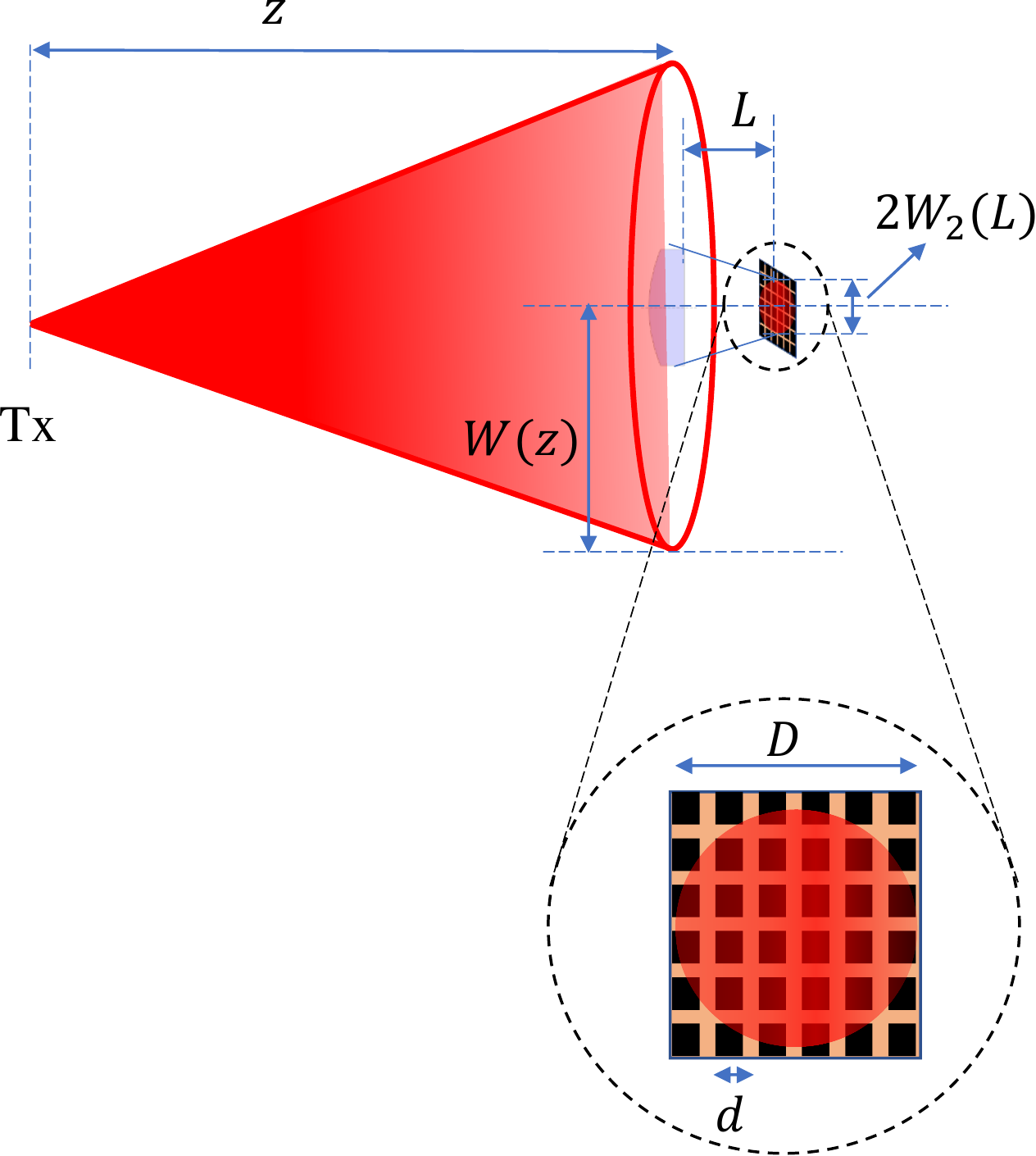}}}
\caption{(a) The proposed $\sqrt{\mathcal{N}_{\rm a}}\times\sqrt{\mathcal{N}_{\rm a}}$ array of arrays structure for imaging receiver, (b) geometry of beam spots and one array of PDs.}
\label{Fig-geoReceiver}
\vspace{-0.6cm}
\end{figure}

Fig.~\ref{sub1:Receiver} illustrates the the proposed array of arrays structure, in which $\mathcal{N}_{\rm a}$ arrays are assumed to be arranged on a square lattice. It can be thus expressed as:\vspace{-0.1cm}
\begin{equation}
    \mathcal{N}_{\rm a}=\left(\frac{D_{\rm a}}{2r_{\rm lns}} \right)^2,\vspace{-0.1cm}
\end{equation}
where $D_{\rm a}$ is the side length of the array of array receiver. 
In this structure, to produce a high received optical power, the combined output of the array of $\mathcal{N}_{\rm a}$ imaging detectors are considered where each imaging detector includes an $N_{\rm PD}$ photodetector array equipped with a collecting lens placed at a distance of $L$ from it (see Fig.~\ref{sub2:Receiver}). 
Note that each array of PDs are assumed to be integrated into a single chip. Let the side length of each \ac{PD} and the array be denoted by $d$ and $D$, respectively. The \ac{FF} of an array of $N_{\rm PD}$ PDs, which are arranged in a square lattice, can be described as:
%
\begin{equation}
    {\rm{FF}}=\frac{ N_{\rm PD} d^2}{D^2} \raisepunct{.}
    \label{Eq-fillfactor}
\end{equation}

\noindent Assuming that the overall size of the receiver is fixed, the size of the outer array (i.e., $\mathcal{N}_{\rm a}$) controls the gain-\ac{FOV} trade-off while the size of the inner arrays (i.e., $N_{\rm PD}$) controls the bandwidth-area trade-off. Therefore the sizes of inner and outer arrays can be adjusted to achieve desired receiver performance in terms of data rate, \ac{FOV}, and \ac{BER}. Fixing the size of arrays, the key remaining control parameters are the side length of each \ac{PD}, $d$, and the distance between the lens and PD array, $L$. Note that defocusing (i.e., reducing $L$ below the focal length) can further increase the \ac{FOV} of the receiver (see Fig.~\ref{sub2:Receiver}). In the following sections, we will initially present the \ac{SNR} analysis of an array of arrays and then, the problem formulation for an optimized design will be introduced.

%

\section{SNR Analysis of the Array of Arrays Receiver}
\label{SNRAnalysis}
We assume that a lens of radius $r_{\rm lns}$ is placed in front of the array, as shown in Fig.~\ref{sub2:Receiver}. It is also assumed that each PD in the array is individually connected to a high bandwidth \ac{TIA}, since the output current of \ac{PIN} photodetectors are typically small. Under this assumption, the receiver bandwidth is limited by the \ac{PD}'s bandwidth.

Three types of noise can affect the performance of this receiver; thermal, shot and \ac{RIN} noise \cite{Sarbazi2000PIMRC, Kazemi2000PIMRC}. The dominant noise source is the TIA thermal noise whose variance is given by \cite{StephenBAlexander1997, Kahn97WirelessInfrared}:
\begin{equation}
\label{NoiseVariance}
    \sigma_{\rm n}^2=\dfrac{4k_{\rm b}TF_{\rm n}B}{R_{\rm f}} \raisepunct{,}
\end{equation}
where $k_{\rm b}$ is the Boltzmann constant, $R_{\rm f}$ is the feedback resistor of the TIA, $T$ is temperature in Kelvin, $F_{\rm n}$ is the noise figure of the TIA and $B$ is the bandwidth of the PIN photodetector. 

Different combining methods such as \ac{MRC}, \ac{EGC} and \ac{SC} can be used to combine the output of PDs. Typically, MRC outperforms the other two techniques while requiring a more complex hardware circuit. In this paper, we provide the analyses for MRC and EGC, however, they can be extended to SC in a similar way. In the following, we start by deriving the \ac{SNR} of the MRC technique for a single array shown in Fig.~\ref{sub1:Receiver}. Next, we will present the SNR derivations of EGC method. We will then extend our analysis to an array of $\sqrt{\mathcal{N}_{\rm a}}\times \sqrt{\mathcal{N}_{\rm a}}$ PD arrays as illustrated in Fig.~\ref{sub2:Receiver}.

The \ac{SNR} of MRC technique for a single array can be obtained as \cite{goldsmith2005wireless}:
%
\begin{equation}
\label{SNR_SingleArray}
    \gamma_{\rm MRC}= \sum_{i=1}^{N_{\rm PD}}\gamma_i,
\end{equation}
where $\gamma_i$ is the SNR of the $i$th PD, which is:
\vspace{-0.2cm}
\begin{equation}
\label{SNRithPD}
    \gamma_i=\dfrac{\left(R_{\rm res}P_{{\rm r},i} \right)^2}{\sigma_{\rm n}^2} \raisepunct{,}\vspace{-0.1cm}
\end{equation}
where $R_{\rm res}$ is the PD responsivity, $\sigma_{\rm n}^2$ is the variance of noise given in \eqref{NoiseVariance}, and $P_{{\rm r},i}$ denotes the received optical power of the $i$th PD. $P_{{\rm r},i}$ depends on the beam spot radius after the lens, $W_2(L)$, which is expressed by \eqref{Eq:beamspotAftrLens} and is shown in Fig.~\ref{sub2:Receiver}.

\begin{figure}[t!]
	\centering  
	\subfloat[\label{sub1:Even1}]{%
		\resizebox{.2\linewidth}{!}{\includegraphics[valign=c]{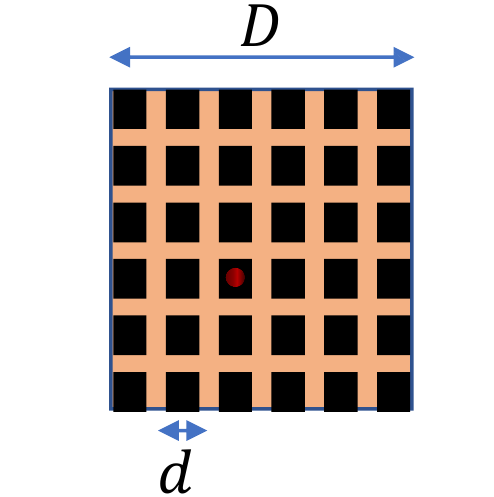}}}\ \
	\subfloat[\label{sub2:Even2}]{%
		\resizebox{.2\linewidth}{!}{\includegraphics[valign=c]{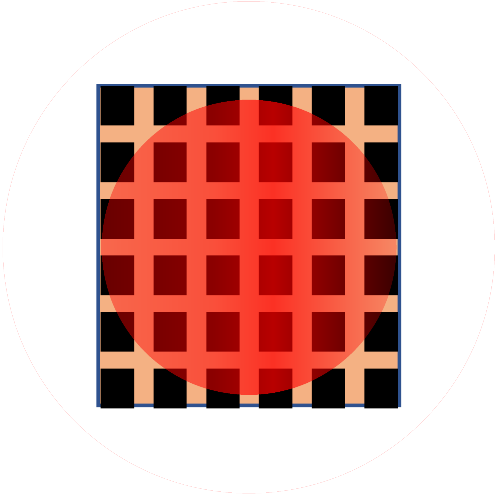}}}\ \
	\subfloat[ \label{sub3:Even3}]{%
		\resizebox{.2\linewidth}{!}{\includegraphics[valign=c]{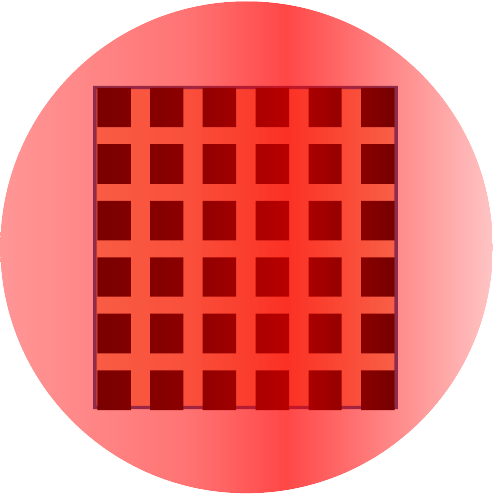}}}\\
		\vspace{-0.1cm}
	\caption{Illustration of different conditions of beam spot radius for an array of PDs.}
	\label{OOandEven}
	\vspace{-0.6cm}
\end{figure}

The radius of beam spot at the receiver plane is denoted by $W(z)$, where it is assumed that $W(z)\gg r_{\rm lns}$. Thus, we can assume almost a uniform beam intensity over the whole area of the lens. Furthermore, ray-tracing simulations conducted in OpticStudio \cite{Zemax} confirm that the beam after the lens has almost a uniform intensity profile. Accordingly, $P_{{\rm r},i}$ can be readily calculated as:
%
\begin{equation}
\label{Eq:PDreceivedPower}
    P_{{\rm r},i}= \iint\limits_{\mathcal{A}_i}\dfrac{\xi P_{{\rm r,lns}}}{\pi W_2^2(L)} \,dx\,dy=\xi P_{{\rm r,lns}}\dfrac{\mathcal{A}_i}{\pi W_2^2(L)},
\end{equation}
where $\mathcal{A}_i$ denotes the overlapping area of beam after the lens and the $i$th PD. Also, $\xi=\xi_{\rm r}\xi_{\rm p}\xi_{\rm a}$; with $\xi_{\rm r}$ representing the transmission coefficient of the lens;  $\xi_{\rm p}=0.5$ due to definition of spot radius being the minimum radius that includes $50\%$ of total power and $\xi_{\rm a}={\rm CA}^2/(2r_{\rm lns})^2$, where ${\rm CA}$ is the clear aperture. 
In \eqref{Eq:PDreceivedPower}, $P_{{\rm r,lns}}$ is the power collected by the lens with the radius of $r_{\rm lns}$.
Based on the relation between $W_2(L)$, $D$ and $d$, there are separate cases for $P_{{\rm r},i}$. 
For example, as shown in Fig.~\ref{sub1:Even1}, 
a small beam spot radius can result in detecting the whole collected power by lens. 
When considering these three possible conditions, the received power by each detector can be expressed as:
%
\begin{equation}
\label{Eq:PDreceivedPowerGeneral}
    P_{{\rm r},i}=
    \begin{cases}
    \xi P_{{\rm r,lns}}\rho_i, \ \ \ \ \ & W_2(L) \leq  \dfrac{1}{\sqrt{\pi}}d,\\
    \xi P_{{\rm r,lns}}\dfrac{\mathcal{A}_i}{\pi W_2^2(L)}, \ \ \ \ \ & \dfrac{1}{\sqrt{\pi}}d < W_2(L) \leq  \dfrac{1}{\sqrt{\pi}}D,\\
    \xi P_{{\rm r,lns}}\dfrac{d^2}{\pi W_2^2(L)}, \ \ \ \ \ & W_2(L) >  \dfrac{1}{\sqrt{\pi}}D.
    \end{cases}
\end{equation}
%
where $\rho_i=1$ if the beam spot after lens hits a detector and $\rho_i=0$ if it does not. Note that, for simplification, in the case of $W_2(L)\leq\frac{1}{\sqrt{\pi}}d$, which is equivalent of a small beam spot at the detector plane (see Fig.~\ref{sub1:Even1}), we approximate the received power by assuming that the whole beam is either detected by a single PD or not detected at all. Moreover, in the case of $W_2(L) >  \frac{1}{\sqrt{\pi}}D$, where the beam spot covers the whole array (see Fig.~\ref{sub3:Even3}), each detector would be fully illuminated thereby $\mathcal{A}_i=d^2$. Substituting \eqref{Eq:PDreceivedPowerGeneral} into \eqref{SNRithPD} and then into \eqref{SNR_SingleArray}, the SNR of MRC technique for a single array is obtained as:
%
\begin{equation}
\label{Eq:SNR-singleArrayAll}
    \gamma_{\rm MRC}=\dfrac{1}{\dfrac{4k_{\rm b}TF_{\rm n}B}{R_{\rm f}}}\times
    \begin{cases}
    \rho\left(R_{\rm res}\xi P_{{\rm r,lns}} \right)^2, \ \ \ \ \ & W_2(L) \leq  \dfrac{1}{\sqrt{\pi}}d,\\
    \left(\dfrac{R_{\rm res}\xi P_{{\rm r,lns}}}{\pi W_2^2(L)} \right)^2\sum\limits_{i=1}^{N_{\rm PD}}\mathcal{A}_i^2, \ \ \ \ \ & \dfrac{1}{\sqrt{\pi}}d < W_2(L) \leq  \dfrac{1}{\sqrt{\pi}}D,\\
    N_{\rm PD}\left(R_{\rm res}\xi P_{{\rm r,lns}}\dfrac{d^2}{\pi W_2^2(L)} \right)^2, \ \ \ \ \ & W_2(L) >  \dfrac{1}{\sqrt{\pi}}D . 
    \end{cases}
\end{equation}
where $\rho=1$ if the beam spot after lens hits one of the detectors in the array and $\rho=0$ if it does not. Let us now calculate the mean SNR at the output of the proposed array of arrays receiver illustrated in Fig.~\ref{sub1:Receiver} by averaging over different received angle of arrivals of the beam within the \ac{FOV} of the receiver. The change of the beam angle of arrival will be equivalent to the shift of beam spot on the PD array plane. This means that in the case of large beam spot, $W_2(L) >  \frac{1}{\sqrt{\pi}}D$, the detected power does not vary significantly by changing the angle of arrival of the beam and therefore the mean SNR would be approximately equal to the SNR given in \eqref{Eq:SNR-singleArrayAll}. Furthermore, in the case of intermediate beam size (i.e., $\frac{1}{\sqrt{\pi}}d < W_2(L) \leq  \frac{1}{\sqrt{\pi}}D$), we can approximate the mean value of $\sum\limits_{i=1}^{N_{\rm PD}}\mathcal{A}_i^2$ averaged over random shifts of beam spot as  
\begin{equation}
\label{SumAi}
    E\left[\sum\limits_{i=1}^{N_{\rm PD}}\mathcal{A}_i^2\right] \approx N_{\rm act} A^2=\pi W_2^2(L){\rm{FF}} d^2 , 
\end{equation}
where $N_{\rm act}\approx \pi W_2^2(L){\rm{FF}}/d^2$ is the average number of fully illuminated PDs and $A=d^2$ is the area of a single PD. Finally, in the case of a small beam spot (i.e., $W_2(L)\leq\frac{1}{\sqrt{\pi}}d$), the mean SNR can be estimated based on the probability of beam hitting a detector considering its random change of angle of arrival within FOV, which is given by $E[\rho]={\rm FF}$.
%
%
%
%
%
Therefore, the average SNR of MRC can be approximated as follows:
\begin{equation}
\label{SNR_av}
\begin{aligned}
    \overline{\gamma}_{\rm MRC}
    &=\dfrac{\sqrt{\mathcal{N}_{\rm a}}}{\dfrac{4k_{\rm b}TF_{\rm n}B}{R_{\rm f}}}\times
    \begin{cases}
     (R_{\rm res}\xi P_{\rm r,lns})^2{\rm{FF}}, & W_2(L) \leq  \dfrac{1}{\sqrt{\pi}}d, \\
     (R_{\rm res}\xi P_{\rm r,lns})^2\dfrac{d^2}{\pi W_2^2(L)}{\rm{FF}}, & \dfrac{1}{\sqrt{\pi}}d < W_2(L) \leq  \dfrac{1}{\sqrt{\pi}}D,\\
     N_{\rm PD}\left(R_{\rm res}\xi P_{\rm r,lns}\dfrac{d^2}{\pi W_2^2(L)}\right)^2, & W_2(L) >  \dfrac{1}{\sqrt{\pi}}D. 
\end{cases}
\end{aligned}
\end{equation}
Note that the accuracy of the approximated average SNR above will be verified in section \ref{Sec:SimulationResults}.
Now, consider the SNR of EGC, which is given by \cite{ZheAngle2014}: 
\begin{equation}
     \gamma_{\rm EGC}=\dfrac{\left(\sum\limits_{i=1}^{N_{\rm PD}}{R_{\rm res}P_{{\rm r},i}}\right)^2}{N_{\rm PD}\sigma_{\rm n}^2}. 
\end{equation}
\noindent Similarly, the average SNR of EGC over various tilt angles within the \ac{FOV} of the receiver for an array of arrays of PIN detectors when thermal noise is dominant can be obtained as: 
\begin{equation}
\label{SNR_EGC_av}
\overline{\gamma}_{\rm EGC}=\dfrac{\sqrt{\mathcal{N}_{\rm a}}}{\dfrac{4k_{\rm b}TF_{\rm n}B}{R_{\rm f}}}\times
    \begin{cases}
     \dfrac{1}{N_{\rm PD}}(R_{\rm res}\xi P_{\rm r,lns})^2{\rm{FF}}, & W_2(L) \leq  \dfrac{1}{\sqrt{\pi}}d, \\
     \dfrac{1}{N_{\rm PD}}(R_{\rm res}\xi P_{\rm r,lns}{\rm{FF}})^2, & \dfrac{1}{\sqrt{\pi}}d < W_2(L) \leq  \dfrac{1}{\sqrt{\pi}}D,\\
     N_{\rm PD}\left(R_{\rm res}\xi P_{\rm r,lns}\dfrac{d^2}{\pi W_2^2(L)}\right)^2, & W_2(L) >  \dfrac{1}{\sqrt{\pi}}D. 
\end{cases}
\end{equation}
By comparing \eqref{SNR_av} and \eqref{SNR_EGC_av}, it can be concluded that both MRC and EGC have a similar performance for $W_2(L)\geq \frac{1}{\sqrt{\pi}}D$. However, when $W_2(L)\leq \frac{1}{\sqrt{\pi}}d$, MRC outperforms EGC around:\vspace{-0.1cm}
\begin{equation}
    G=10\log_{10}\left(\frac{\overline{\gamma}_{\rm MRC}}{\overline{\gamma}_{\rm EGC}}\right)=10\log_{10}(N_{\rm PD})\ \ \  {\rm [dB]}.
    \vspace{-0.1cm}
\end{equation}
Larger values of $N_{\rm PD}$ can result in a significant gain difference between MRC and EGC. Due to this notable difference, we will consider MRC for the rest of our analyses. 

\section{Optimum Geometric Design of High-Speed Receiver}
\label{Sec: Optimum Geometric Design}
In this section, assuming a fixed size for inner arrays, we aim to optimize the side length of each \acp{PD} in the array, $d$, as well as the distance between the lens and the array, $L$, to achieve the maximum data rate while satisfying a desired receiver \ac{FOV}.  
In order to develop a compact receiver, $L$ is assumed to be limited to the range $0\leq L\leq f_{\rm b}$. That is, the lens is not placed at a distance more than its focal length from the PD array. This allows for both focused ($L = f_{\rm b}$) and defocused ($L < f_{\rm b}$) designs while the latter can relax the \ac{FOV} constraint to some degree. It will be shown that the optimum value of $L$ can be specified according to the desired \ac{FOV}. In addition, the optimum value of $d$ is selected in such a way that it achieves the maximum data rate while fulfilling the required BER (or SNR). We note that although small PDs have a higher bandwidth, they add more noise to the system thereby degrading the \ac{SNR}. On the other hand, large area PDs decrease the system bandwidth and consequently the achievable data rate. Such a behavior in a power-limited regime yields an optimum $d$ that maximizes the data rate. 
The optimization problem can be formulated as:
\vspace{-0.3cm}
\begin{subequations}
\label{OP1}
\begin{align}   
    \max\limits_{L,d} \ \ \ &R \\
    \rm{s.t.} \ \ \ &{\rm {FOV}}(L)\geq\rm {FOV}_{\rm req},\\\label{24c}
    &\overline{\gamma}_{\rm MRC}(L,d)\geq \gamma_{\rm req},\\
    &d_{\rm min}\leq d\leq d_{\rm max}=D\sqrt{\frac{{\rm{FF}}}{N_{\rm PD}}}. 
\end{align}
\end{subequations}
The first constraint guarantees the FOV requirement and leads to an upper bound on $L$ as $L\leq L_{\rm max}$, where $L_{\rm max}$ can be obtained by substituting $\rm {FOV}_{\rm req}$ in \eqref{Eq:L-FOV} as:
\begin{equation}
\label{LandFOV_req}
    L_{\rm max}=a_3 {\rm FOV}_{\rm req}^3+a_2 {\rm FOV}_{\rm req}^2+a_1 {\rm FOV}_{\rm req}+a_0. \vspace{-0.2cm}
\end{equation} The second constraint enforces limitations on both the size of PDs and the distance between the lens and the array. In fact, SNR is directly related to $d$ and $L$, where small values of $d$ and $L$ result in lower SNRs. By substituting \eqref{Bandwidth-SimplifiedEq} and \eqref{Eq-fillfactor} into \eqref{SNR_av}, we can express this relationship as:\vspace{-0.1cm}
\begin{equation}
    \label{simplified-Av:SNR}
    \begin{aligned}
    \overline{\gamma}_{\rm MRC}
    =\frac{1}{A_x}\times
    \begin{cases}
     d^3, & W_2(L) \leq  \dfrac{1}{\sqrt{\pi}}d, \\
     \dfrac{d^5}{\pi W_2^2(L)}, & \dfrac{1}{\sqrt{\pi}}d < W_2(L) \leq  \dfrac{1}{\sqrt{\pi}}D,\\
     D^2\dfrac{d^5}{\left(\pi W_2^2(L)\right)^2}, & W_2(L) >  \dfrac{1}{\sqrt{\pi}}D, \vspace{-0.1cm}
\end{cases}
\end{aligned}
\end{equation}
where
%
\begin{equation}
    A_x = \dfrac{4k_{\rm b}TF_{\rm n} D^2}{N_{\rm PD}\sqrt{\mathcal{N}_{\rm a}} C_{\rm t} R_{\rm f} (R_{\rm res}\xi P_{\rm r,lns})^2 } \raisepunct{.} 
\end{equation}
In constraint (\ref{24c}), $\gamma_{\rm req}$ is the required SNR to ensure that BER is less than a target BER, i.e., $\rm {BER}_{\rm req}$.
The threshold for SNR can be represented based on the BER requirement as \cite{ghassemlooy2019optical}:
%
\begin{equation}
    \gamma_{\rm req}=\left( Q^{-1}(\rm BER_{req})  \right)^2, 
\end{equation}
where $Q^{-1}(\cdot)$ is the inverse of $Q$-function.
The size of each PD on the array is limited to $d_{\rm min}$ and $D\sqrt{\frac{{\rm{FF}}}{N_{\rm PD}}}$. The right side of the third constraint is obtained based on \eqref{Eq-fillfactor} and ensures the fill factor of the array is limited to ${\rm{FF}}$.  
To guarantee that the designed inner arrays follow a square lattice configuration, the number of PDs on each array are selected from the set of square numbers. The transmit power in this optimization problem is set to the maximum permissible value according to the eye safety regulations, i.e., $P_{\rm t}=P_{\rm t,max}$. The details of the eye safety considerations for obtaining $P_{\rm t,max}$ can be found in \cite{MDSoltani2021safety}. 
The optimization problem in \eqref{OP1} along with the piecewise SNR function in \eqref{simplified-Av:SNR} can be broken down into three separate optimization problems, where the global solution will be the optimal solution among the three possible solutions. These optimization problems are expressed below.

\noindent Optimization problem 1: \vspace{-0.3cm}
\begin{subequations}
\label{OP1-General}
\begin{align}   
    \max\limits_{d,L} \ \ \ &R \\ \vspace{-0.1cm}
    \rm{s.t.} \ \ \ &{\rm {FOV}}(L)\geq\rm {FOV}_{\rm req},\\ \vspace{-0.1cm}
    &\frac{d^3}{A_x}\geq \gamma_{\rm req},\\ \vspace{-0.1cm}
    &d_{\rm min}\leq d\leq D\sqrt{\frac{{\rm{FF}}}{N_{\rm PD}}}, \\ \vspace{-0.1cm}
    &d\geq \sqrt{\pi}W_2(L). \label{OP1-General-const4}\vspace{-0.3cm}
\end{align}
\end{subequations}

\noindent Optimization problem 2: \vspace{-0.3cm}
\begin{subequations}
\label{OP2-General}
\begin{align}   
    \max\limits_{d,L} \ \ \ &R \\
    \rm{s.t.} \ \ \ &{\rm {FOV}}(L)\geq\rm {FOV}_{\rm req},\\
    &\frac{d^5}{A_x\pi W_2^2(L)}\geq \gamma_{\rm req},\\
    &d_{\rm min}\leq d\leq D\sqrt{\frac{{\rm{FF}}}{N_{\rm PD}}}, \\
    &W_2(L)>\frac{d}{\sqrt{\pi}},\\ \label{OP2-General-const3}
    &W_2(L)\leq \dfrac{1}{\sqrt{\pi}}D,\vspace{-0.2cm}
\end{align}
\end{subequations}

\noindent Optimization problem 3: \vspace{-0.3cm}
\begin{subequations}
\label{OP3-General}
\begin{align}   
    \max\limits_{d,L} \ \ \ &R \\
    \rm{s.t.} \ \ \ &{\rm {FOV}}(L)\geq\rm {FOV}_{\rm req},\\
    &\frac{d^5}{A_x\left(\pi W_2^2(L)\right)^2}D^2\geq \gamma_{\rm req},\\ 
    &d_{\rm min}\leq d\leq D\sqrt{\frac{{\rm{FF}}}{N_{\rm PD}}}, \\
    &W_2(L)> \dfrac{1}{\sqrt{\pi}}D,\vspace{-0.2cm}
\end{align}
\end{subequations}

In what follows, we solve these optimization problems for two well-known and practical modulation schemes; DCO-OFDM with adaptive QAM and OOK.

\subsection{DCO-OFDM}
\label{Sec:Variable Rate QAM}
To ensure a high spectral efficiency with intensity modulation and direct detection, we here implement the \ac{DC}-biased optical OFDM with adaptive QAM. By properly choosing the variance of the OFDM signal and the \ac{DC} bias, a tight upper bound for the achievable data rate is given by \cite{Sarbazi2000PIMRC,Kazemi2000PIMRC}:
\vspace{-0.3cm}
\begin{equation}
\label{VariableRateQAM}
    R=\nu B \log_2\left(1+\frac{\overline{\gamma}_{\rm MRC}}{\Gamma} \right), \vspace{-0.2cm}
\end{equation}
where $\nu=\frac{N_{\rm sc}-2}{N_{\rm sc}}$ with $N_{\rm sc}$ denoting the number of subcarriers. In \eqref{VariableRateQAM}, $\Gamma$ represents the SNR gap due to the required BER, which is given by:
\vspace{-0.1cm}
\begin{equation}
    \Gamma=\frac{-\ln(5{\rm BER_{\rm req}})}{1.5}\raisepunct{.} \vspace{-0.1cm}
\end{equation}
The optimization problem defined in \eqref{OP1} can be applied here to find the optimum side length of PDs and the distance between the lens and the inner array to maximize the data rate. 
Here, $R$ is a differentiable function with respect to (w.r.t) $d$ and $L$, showing different behaviors in different optimization problems discussed above. 
%
%
We now solve the optimization problems given in \eqref{OP1-General}, \eqref{OP2-General} and \eqref{OP3-General} for DCO-OFDM. 

\noindent \textbf{Optimization problem 1:} In this case, the objective function of \eqref{OP1-General} defined in (\ref{VariableRateQAM}) can be expressed as: \vspace{-0.3cm}
%
\begin{equation}
    \label{OP1-const1-OFDM}
    R=\dfrac{\nu}{C_{\rm t}d} \log_2\left(1+\frac{d^3}{\Gamma A_x} \right).
\end{equation}

\begin{figure}[t!]
 \centering
\includegraphics[width=0.55\textwidth]{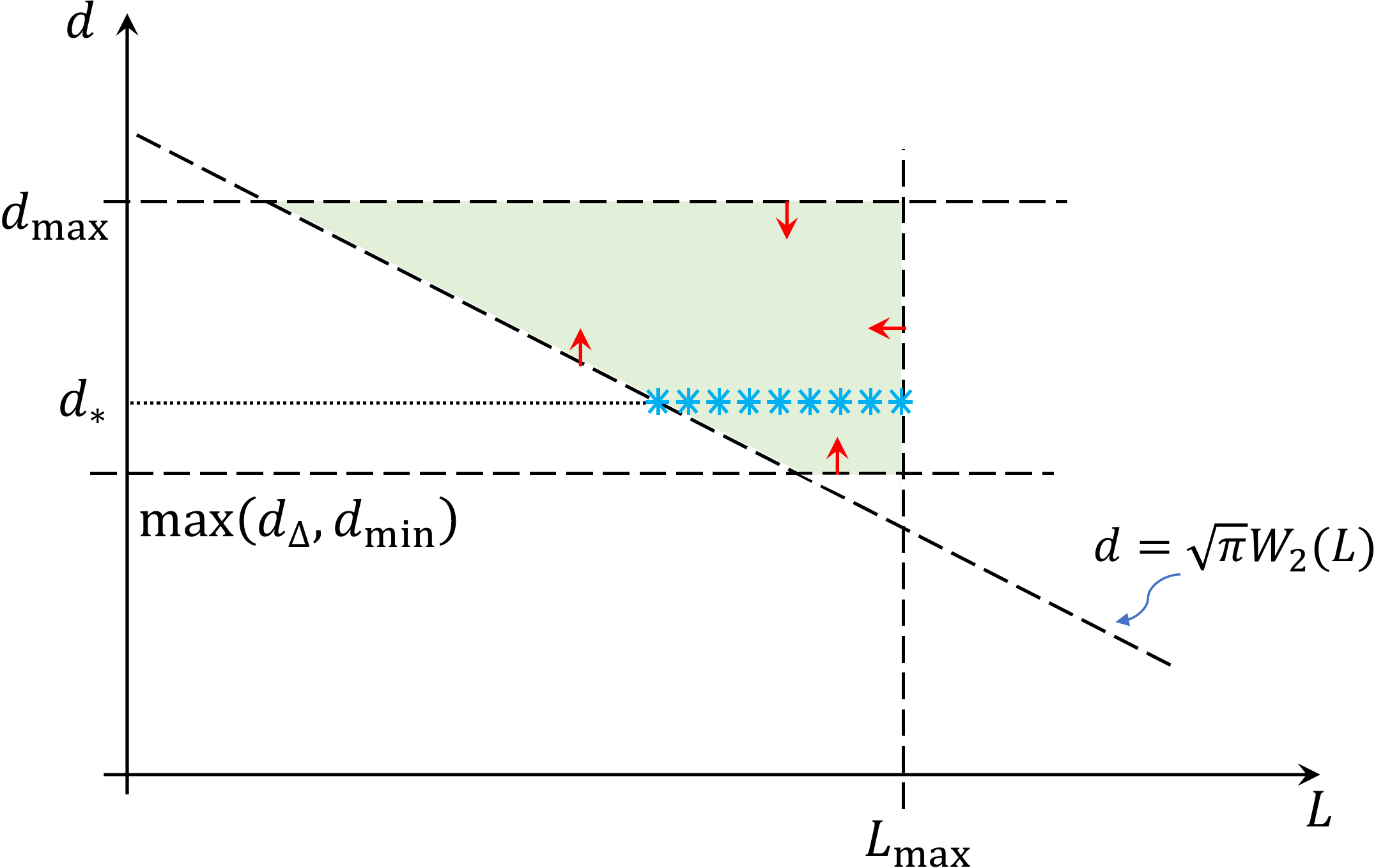} 
\caption{One example of possible solutions for optimization problem given in \eqref{OP1-General}. }
\label{Fig-OFDM-Pr1}
\vspace{-0.5cm}
\end{figure}

The function in \eqref{OP1-const1-OFDM} is only a function of $d$ and can be shown to have a unique maximum at: \vspace{-0.3cm}
\begin{equation}
\label{Root-Derivative-RateOFDM}
    d_*=\left({15.8 \Gamma A_x}\right)^{\frac{1}{3}}. \vspace{-0.2cm}
\end{equation}
The objective function also has a minimum at $d=0$  which is outside the feasible set. 
We should identify whether the possible solution given by (\ref{Root-Derivative-RateOFDM}) is within the feasible region of the problem 1. 
The second constraint of \eqref{OP1-General} implies:\vspace{-0.3cm}
\begin{equation}
\label{SNR-firstFunc}
    d\geq d_{\Delta}\triangleq\left({A_x \gamma_{\rm req}}\right)^{\frac{1}{3}}, \vspace{-0.2cm}
\end{equation}
and the third constraint enforces $d\geq d_{\rm min}$. Therefore, these two apply the condition $d\geq \max(d_{\Delta},d_{\rm min})$. 
Fig.~\ref{Fig-OFDM-Pr1} shows an example of the feasible region of the problem 1 defined by the constraints in \eqref{OP1-General}. 
The red arrows indicate the direction of feasibility for each boundary curve defined by different constraints in \eqref{OP1-General}. For instance, the region of interest for the condition $d\leq d_{\max}$ is shown with the downward arrow. The blue stars show the maximum points of $R$. If $\max(d_{\Delta},d_{\min})<d_*<d_{\max}$ as shown in the figure, the function $R$ is maximized at $d_{\rm opt,1}=d_*$. However, if $d_*<\max(d_{\Delta},d_{\min})$, $R$ is a decreasing function w.r.t $d$ in the feasible region and the optimum $d$ is $d_{\rm opt,1}=\max(d_{\Delta},d_{\min})$.  Otherwise, if $d_*>d_{\max}$, $R$ is an increasing function w.r.t $d$, and therefore, $R$ is maximized at $d_{\rm opt,1}=d_{\max}$. These cases are summarized in Table~\ref{VariableRateSNR1-Table}. As can be seen from Fig.~\ref{Fig-OFDM-Pr1}, the optimum $L$ can be from the cross-section of line $d=\sqrt{\pi}W_2(L)$ and $d=d_{\rm opt,1}$ to $L_{\max}$, i.e., in the range $[\mathcal{L}(d_{\rm opt,1}), L_{\rm max}]$, where $\mathcal{L}(\cdot)$ is given by: \vspace{-0.2cm}
\begin{equation}
\label{LDelta}
    \mathcal{L}(x)=f_{\rm b}-\frac{1}{b_1}\left(\frac{1}{\sqrt{\pi}}x-b_0 \right). 
    \vspace{-0.2cm}
\end{equation} 


\begin{table}[]
\centering
\caption[blah]{Optimum $d$ for the first optimization problem of DCO-OFDM.}
\label{VariableRateSNR1-Table}
\begin{tabular}{|c|c|c|c|}
\hline
Condition        & Optimum $d$  & Optimum $L$  & Behavior of $R(d)$ \\ \hline \hline
$d_*<\max(d_{\Delta},d_{\min})$ & $\max(d_{\Delta},d_{\min})$ & $[\mathcal{L}(\max(d_{\Delta},d_{\min})), L_{\rm max}]$  & decreasing  \\ \hline
$\max(d_{\Delta},d_{\min})<d_*<d_{\max}$ & $d_*$ & $[\mathcal{L}(d_*), L_{\rm max}]$ & \multicolumn{1}{c|}{\begin{tabular}[c]{@{}c@{}}increasing till $d_*$\\  then decreasing\end{tabular}} \\ \hline
$d_*>d_{\max}$ & $d_{\max}$ & $[\mathcal{L}(d_{\max}), L_{\rm max}]$ & increasing \\ \hline
\end{tabular}
\vspace{-0.7cm}
\end{table}

\begin{figure}[t!]
 \centering
\includegraphics[width=0.55\textwidth]{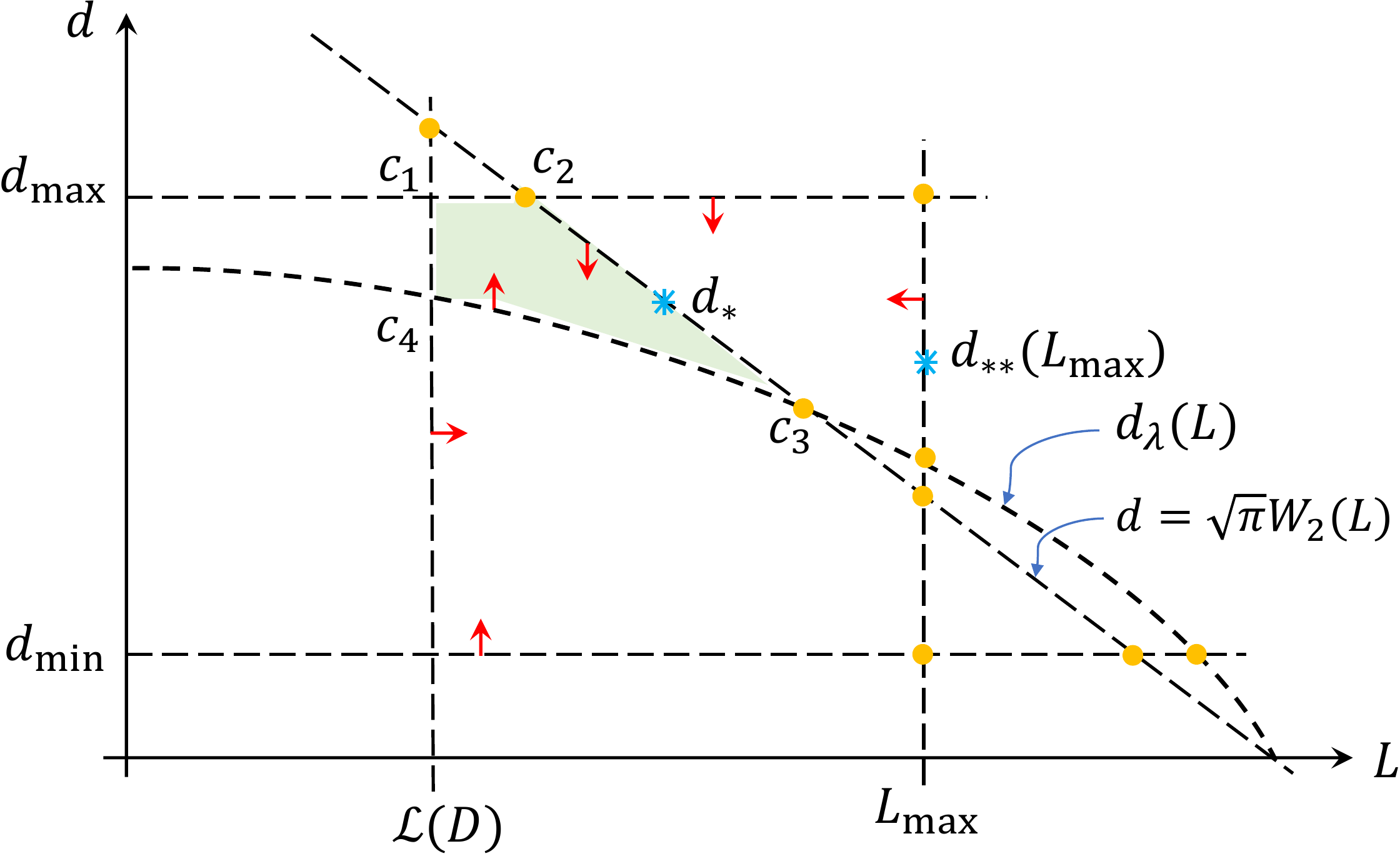} 
\caption{One example of possible solutions for optimization problem given in \eqref{OP2-General}.}
\label{Fig-OFDM-Pr2}
\vspace{-0.5cm}
\end{figure}

\noindent \textbf{Optimization problem 2:} The objective function of \eqref{OP2-General} is given by: \vspace{-0.3cm}
%
%
\begin{equation}
    \label{OP2-OFDM-const1}
    R=\dfrac{\nu}{C_{\rm t}d} \log_2\left(1+\frac{d^5}{\pi W_2^2(L) \Gamma A_x}\right),
\end{equation}
which is a function of both $d$ and $L$. For any $L$, the function in (\ref{OP2-OFDM-const1}) has a unique maximum at: \vspace{-0.2cm}
\begin{equation}
\label{extreme-cond-SNR2}
    d_{**}(L)=\left({142.32  \pi W_2^2(L) \Gamma A_x} \right)^{\frac{1}{5}}. \vspace{-0.2cm}
\end{equation}
Moreover, the objective function in (\ref{OP2-OFDM-const1}) is an increasing function w.r.t $L$, for any $d$. This implies that the solution should be always on the boundary of the feasible region since for any internal point, there exists a point on the boundary of the feasible region with the same $d$ and a higher value of $L$, which results in a higher $R$. Fig.~\ref{Fig-OFDM-Pr2} represents one example of the feasible region of the problem 2 imposed by the constraints of \eqref{OP2-General}.
The possible solutions would be either one of the corner points (shown as yellow circles) or a extremum point that maximize $R$ on a boundary curve (shown as blue stars). The curve $d_{\lambda}(L)$ shows the boundary of the second constraint of \eqref{OP2-General} where: \vspace{-0.3cm}
\begin{equation}
\label{SNR-secondFunc}
    d_{\lambda}(L)\triangleq \left( \pi W_2^2(L) A_x \gamma_{\rm req} \right)^{\frac{1}{5}}. \vspace{-0.2cm}
\end{equation}
Note that inserting this curve in the objective function makes $R$ to become independent of $L$ and decreasing w.r.t $d$ with no extremum point. By inserting the fourth constraint of (\ref{OP2-General}) (i.e., $W_2(L)=\frac{d}{\sqrt{\pi}}$) in (\ref{OP2-OFDM-const1}), the objective function simplifies into the objective function of problem 1 given by (\ref{OP1-const1-OFDM}) and therefore, it can be maximized at $d_*$ as in (\ref{Root-Derivative-RateOFDM}). In addition, the 5th constraint of (\ref{OP2-General}) imposes the lower bound $L\geq \mathcal{L}(D)$. Note that most of the corner points on this boundary line do not produce a possible solution. There would exist points with larger $L$ in the feasible region. For the example shown in Fig.~\ref{Fig-OFDM-Pr2}, $d_*$ is the optimum solution. If $d_*$ was out of the feasible region, then $c_2$ (or $c_3$) may be the optimum solution if $R$ is an increasing (or decreasing) function. 
The possible corner and extremum points are given in Table~\ref{Table:CornerPointsPr2}. Therefore, the solution of problem 2 can be determined by finding the optimum point among these possible solutions that satisfy all the constraints.  



\begin{table}[]
\centering
\caption[blah]{Possible optimum solutions for $(d,L)$ for the second optimization problem of DCO-OFDM.}
\label{Table:CornerPointsPr2}
\begin{tabular}{|c|c|c|}
\hline
\multicolumn{3}{|c|}{Possible optimum solutions for $(d,L)$}     \\ \hline                                                                                                 
$(d_{\max},L_{\max})$                                  & $(d_{\min},\mathcal{L}(d_{\min} ))$                                          & $(D,\mathcal{L}(D))$ \\ \hline
$(d_{\lambda}(L_{\max}),L_{\max})$                     & $(d_{\min},\mathcal{L}(\sqrt{\frac{d_{\min}^5}{A_x \gamma_{\rm req}}})$      & $(d_*,\mathcal{L}(d_*))$ \\ \hline
$(\sqrt{\pi}W_2(L_{\max}),L_{\max})$                   & $(d_{\Delta},\mathcal{L}(d_{\Delta}))$                                       & $(d_{**}(L_{\max}),L_{\max})$  \\ \hline
$(d_{\min},L_{\max})$                                  & $(d_{\max},\mathcal{L}(d_{\max} ))$                                          &  -  \\ \hline
\end{tabular}
\vspace{-0.8cm}
\end{table}

\begin{figure}[t!]
 \centering
\includegraphics[width=0.55\textwidth]{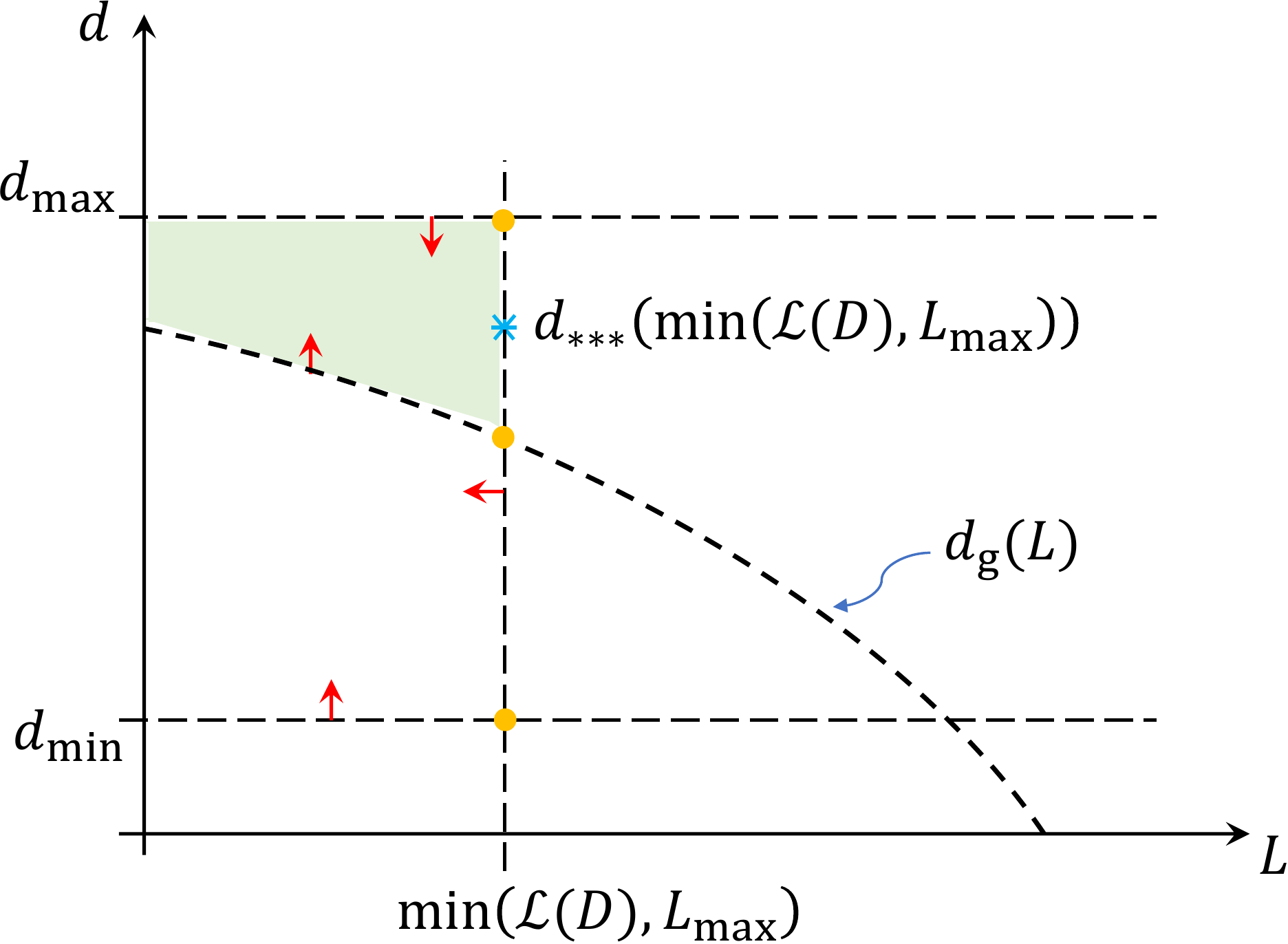} 
\caption{One example of possible solutions for optimization problem given in \eqref{OP3-General}.}
\label{Fig-OFDM-Pr3}
\vspace{-0.5cm}
\end{figure}

\noindent \textbf{Optimization problem 3:} The objective function of \eqref{OP3-General} can be expressed as:
%
%
\begin{equation}
    \label{Rate:OFDM-Pro3}
    R=\dfrac{\nu}{C_{\rm t}d} \log_2\left(1+\frac{D^2 d^5}{\pi^2 W_4^2(L) \Gamma A_x}\right).
\end{equation}
This function has a similar behavior to the objective function of problem 2 in (\ref{OP2-OFDM-const1}) with a unique maximum point for any $L$ at: \vspace{-0.2cm}
\begin{equation}
\label{extreme-cond-SNR3}
    d_{***}(L)=\left(\frac{142.32 \pi^2 W_2^4(L) \Gamma A_x }{D^2} \right)^{\frac{1}{5}}. \vspace{-0.1cm}
\end{equation}
The objective function is also increasing w.r.t to $L$ for any $d$. Therefore, the solution lies again on the boundary of the feasible region. Fig.~\ref{Fig-OFDM-Pr3} illustrates one example of the feasible region imposed by the constraints of \eqref{OP3-General}. The curve labeled as $d_{\rm g}(L)$ is the boundary condition imposed by the second constraint in \eqref{OP3-General} where: \vspace{-0.2cm}
\begin{equation}
\label{SNR-thirdFunc}
    d_{\rm g}(L)\triangleq \left(\pi^2 W_2^4(L) \dfrac{A_x}{D^2} \gamma_{\rm req} \right)^{\frac{1}{5}}.  \vspace{-0.2cm}
\end{equation}
%
Note that inserting $d_{\rm g}(L)$ in the objective function (\ref{Rate:OFDM-Pro3}), makes $R$ a decreasing function of $d$ and independent of $L$ with no extremum. In addition, based on the first and fourth constraints of \eqref{OP3-General}, $L\leq \min(\mathcal{L}(D),L_{\rm max})$. 
Therefore, the possible solutions are either the extremum or the corner points on the upper boundary of $L$ as shown in Fig.~\ref{Fig-OFDM-Pr3}. Table~\ref{VariableRateSNR3-Table} summarizes different conditions under which any of these points would be the solution of problem 3. 

The global optimal solution of DCO-OFDM can be then obtained based on the optimum solution of each optimization problem, i.e., the one that results in the maximum data rate. 


\begin{table}[]
\centering
\caption[blah]{Optimum $d$ for the third optimization problem of DCO-OFDM. The optimum $L$ is $\min(\mathcal{L}(D),L_{\max})$ for all optimum $d$'s.}
\label{VariableRateSNR3-Table}
\resizebox{\columnwidth}{!}{%
\begin{tabular}{|c|c|}
\hline
Condition        & Optimum $d$     \\ \hline \hline
$d_{***}(\min(\mathcal{L}(D),L_{\max}))<\max(d_{\min},d_{\rm g}(\min(\mathcal{L}(D),L_{\max})))$ & \!\!$\max(d_{\min},d_{\rm g}(\min(\mathcal{L}(D),L_{\max})))$  \\ \hline
\!\!$\max(d_{\min},d_{\rm g}(\min(\mathcal{L}(D),L_{\max})))<d_{***}(\min(\mathcal{L}(D),L_{\max}))<d_{\max}$\! & $d_{***}(\min(\mathcal{L}(D),L_{\max}))$  \\ \hline
$d_{***}(\min(\mathcal{L}(D),L_{\max}))>d_{\max}$ & $d_{\max}$ \\ \hline
\end{tabular}
}
\vspace{-0.5cm}
\end{table}

\vspace{-0.2cm}
\subsection{OOK Modulation}
\vspace{-0.2cm}
\label{Sec:OOK-modulation}
The data rate of OOK modulation is limited by the Nyquist's sampling rate, i.e., $R\leq 2B$. Although this requires employing efficient pulse shaping techniques, experimental studies have shown that data rates close to the Nyquist limit are achievable for OOK modulation \cite{Minh2009OOK,Watson:13-OOK,Lee:15-OOK}. 
The linear relationship between $R$ and $B$, indicates that $B$ can replace $R$ as the objective function of the optimization problem in \eqref{OP1-General}, \eqref{OP2-General} and \eqref{OP3-General} for OOK modulation. Furthermore, replacing \eqref{Bandwidth-SimplifiedEq} for $B$, we have:\vspace{-0.2cm}
\begin{equation}
    \label{Rate:OOK}
    R=\frac{2}{C_{\rm t}d}.
    \vspace{-0.2cm}
\end{equation}
Note that the objective function is a monotonically decreasing function of $d$ and does not depend on $L$; hence the maximized solution should happen at the boundary of the feasible set, i.e.,  at the minimum value of $d$. 
Next, we find the optimum (i.e., minimum) $d$ within the feasible set of the three optimization problem given in \eqref{OP1-General}-\eqref{OP3-General}.


\noindent \textbf{Optimization problem 1:} 
Similar to the first optimization problem of DCO-OFDM, the second and third constraints of \eqref{OP1-General} implies $d\geq \max(d_{\Delta},d_{\min})$, where $d_{\Delta}$ is defined in \eqref{SNR-firstFunc}.
Since $W_2(L)$ is a decreasing function of $L$, inserting $L=L_{\rm max}$ in \eqref{OP1-General-const4} gives a lower limit on the value of $d$. Therefore, considering the lower bounds defined by the second, third, and fourth constraints in \eqref{OP1-General}, the optimum (i.e., minimum) value of $d$ can be expressed as:   \vspace{-0.2cm}
\begin{equation}
\label{conditionI}
d_{\rm opt,1}=\max\left(d_{\rm min},d_{\Delta},\sqrt{\pi}W_2(L_{\rm max})\right), \vspace{-0.2cm}
\end{equation}
for $\max\left(d_{\Delta},\sqrt{\pi}W_2(L_{\rm max})\right)\leq D\sqrt{\frac{{\rm{FF}}}{N_{\rm PD}}}$; otherwise this condition does not provide a solution.
Also note that $L$ should be taken from $[\mathcal{L}(d_{\rm opt,1}), L_{\rm max}]$. 

%

\noindent \textbf{Optimization problem 2:} 
The second constraint of \eqref{OP2-General} indicates $d\geq d_{\lambda}(L)$, where $d_{\lambda}(L)$ is given in \eqref{SNR-secondFunc}.
Noting that $d_{\lambda}(L)$ is a decreasing function of $L$, the minimum value of $d$ (to ensure maximum data rate) can be obtained on the boundary of feasible set at the maximum possible value of $L$. The maximum $L$ may be either on the line $L=L_{\max}$ or on the cross-section of the lines $d=\sqrt{\pi}W_2(L)$ and $d_{\lambda}(L)$ (see Fig.~\ref{Fig-OFDM-Pr2}).  It can be shown that the cross-section of the two lines happens at $L=\mathcal{L}((A_x \gamma_{\rm req} )^{\frac{1}{3}})$. Therefore, the optimum value of $L$ is $\min(\mathcal{L}((A_x \gamma_{\rm req} )^{\frac{1}{3}})),L_{\max})$.  
Applying the third constraint in \eqref{OP2-General}, the solution for $d$ is given by: \vspace{-0.3cm}
\begin{equation}
\label{conditionII}
    d_{\rm opt,2}=\max\left(d_{\rm min},d_{\lambda}(\min(\mathcal{L}((A_x \gamma_{\rm req} )^{\frac{1}{3}})),L_{\max})\right),  \vspace{-0.2cm}
\end{equation}
for $d_{\lambda,\min}\leq D\sqrt{\frac{{\rm{FF}}}{N_{\rm PD}}}$; otherwise this condition does not provide a solution.
%

%

\noindent \textbf{Optimization problem 3:} 
The second constraint of this optimization problem gives $d\geq d_{\rm g}(L)$ where $d_{\rm g}(L)$ is defined in \eqref{SNR-thirdFunc}.
%
Based on the first and fourth constraints of \eqref{OP3-General}, $L\leq \min(\mathcal{L}(D),L_{\rm max})$. Similarly, $d_{\rm g}(L)$ is a decreasing function of $L$; hence, the minimum value of $d$ is achieved on the boundary of feasible set at the maximum value of $L$. Similarly, by applying the third constraint in \eqref{OP3-General}, the solution for $d$ is obtained as:
\vspace{-0.3cm}
\begin{equation}
\label{conditionIII}
    d_{\rm opt,3}=\max\left(d_{\rm min},d_{\rm g}(\min(\mathcal{L}(D),L_{\rm max}))\right).  \vspace{-0.2cm}
\end{equation}
for $d_{\rm g}(\min(\mathcal{L}(D),L_{\rm max}))\leq D\sqrt{\frac{{\rm{FF}}}{N_{\rm PD}}}$; otherwise this condition does not provide a solution.

The solutions of these three optimization problems are outlined in Table~\ref{OOK-OP-Solution}. The global optimal solution of OOK can be then obtained based on the minimum value of \eqref{conditionI}, \eqref{conditionII} and \eqref{conditionIII}. 


\begin{table}[]
\centering
\caption[blah]{The possible solutions of the optimization problem for OOK.}
\label{OOK-OP-Solution}
\begin{tabular}{|c|c|}
\hline
Optimum $L$               & Optimum $d$    \\ \hline \hline
$[L_{\Delta}, L_{\rm max}]$ & $d_{\rm opt,1}=\max\left(d_{\rm min},d_{\Delta},\sqrt{\pi}W_2(L)\right)$ \\ \hline
$\min(\mathcal{L}((A_x \gamma_{\rm req} )^{\frac{1}{3}})),L_{\max})$ & $d_{\rm opt,2}=\max\left(d_{\rm min},d_{\lambda}(\min(\mathcal{L}((A_x \gamma_{\rm req} )^{\frac{1}{3}})),L_{\max})))\right)$                                  \\ \hline      
$\min(\mathcal{L}(D),L_{\rm max})$     &   $d_{\rm opt,3}=\max\left(d_{\rm min},d_{\rm g}(\min(\mathcal{L}(D),L_{\rm max}))\right)$                        \\  \hline
\end{tabular}
\vspace{-0.4cm}
\end{table}

\vspace{-0.2cm}
\section{Simulation Results}
\label{Sec:SimulationResults}
A single mode \ac{VCSEL} transmitter is considered which operates at $850$ nm with a beam waist of $5~\mu$m. A lens is placed in front of the VCSEL source to produce a $20$ cm beam spot diameter at $2$ m. The transmit power of $10$ mW is obtained based on the eye safety considerations \cite{MDSoltani2021safety}, which enforces a power-limited regime.
The maximum size of receiver is set to be $2$ cm $\times$ $2$ cm. The \ac{FF} of the array is set to ${\rm{FF}}=0.64$ similar to the fabricated array in \cite{Umezawa2022LargeSubmil}. 
We have fixed the size of each inner array to $400$ $\mu$m $\times$ $400$ $\mu$m (similar to \cite{Umezawa2022LargeSubmil}). Each inner array is equipped with the aspheric lens $354140-$B from Thorlabs\footnote{This lens is just an example and other lenses can be also used.}.
The lens parameters are provided in Table~\ref{ThorlabAsphericTable}. We have also evaluated the transmission coefficient of the considered aspheric lens using OpticStudio by calculating the ratio between the transmission power through the lens and its incident power. The results show that the transmission coefficient is $\xi_{\rm r}=0.88$. The rest of the simulation parameters are provided in Table~\ref{SimulationParameters}.

\begin{table}[]
\small
\centering
\caption[blah]{Simulation Parameters.}
\label{SimulationParameters}
\begin{tabular}{|c|c|c|}
\hline
\renewcommand{\arraystretch}{1.5}
{Parameter}         & {Symbol}      & {Value}     \\  \hline \hline
Feedback resistor & $R_{\rm f}$ & $500$ $\Omega$ \\  \hline
Load resistor & $R_{\rm L}$ & $50$ $\Omega$ \\  \hline
Junction series resistor & $R_{\rm s}$ & $7$ $\Omega$ \cite{Dupuis2014Exploring} \\  \hline
Carrier saturation velocity & $v_{\rm s}$ & $4.8\times 10^4$ m/s\\ \hline
TIA noise figure  & $F_{\rm n}$ & $5$ dB  \\ \hline
Temperature       & $T$         & $300^{\circ}$ K \\ \hline
Side length of inner array        & $D$         & $400\ \mu$m  \\ \hline 
Transmission coefficient   & $\xi_{\rm r}$ & $0.88$  \\ \hline
Transmit power    & $P_{\rm t}$     & $10$ mW \\ \hline
Receiver responsivity & $R_{\rm res}$ & $0.5$ A$/$W\\ \hline
\end{tabular}
\vspace{-0.3cm}
\end{table}


\begin{figure}[t!]
 \centering
\includegraphics[width=0.5\textwidth]{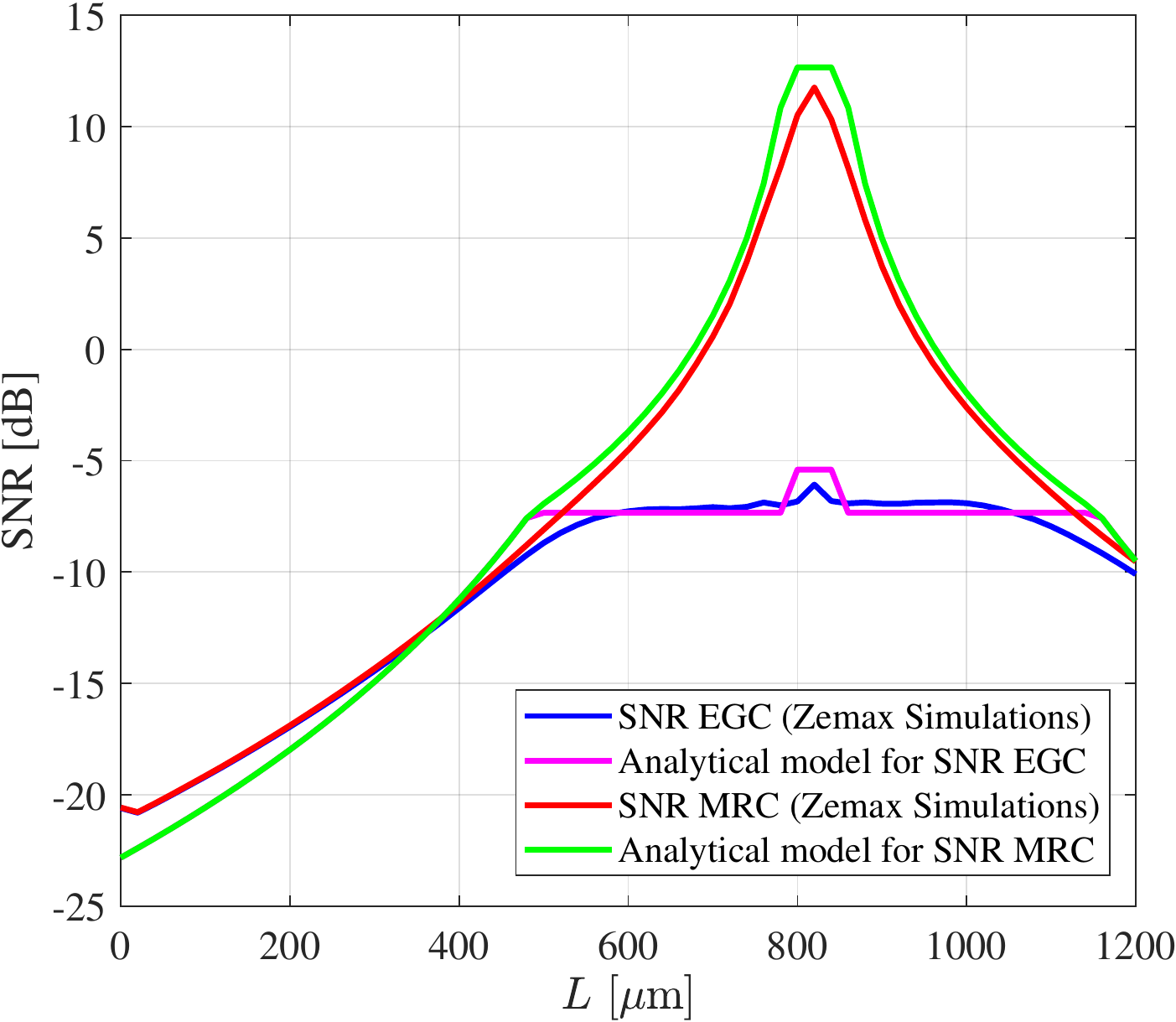} 
\caption{SNR versus $L$ for EGC, MRC and the analytical model based on \eqref{SNR_av} for $\mathcal{N}_{\rm a}=64$ and $N_{\rm PD}=64$.}
\label{Fig-SNR_MRC}
\vspace{-0.8cm}
\end{figure}

Fig.~\ref{Fig-SNR_MRC} show the SNR results of MRC and EGC obtained from OpticStudio. The results are shown for $\mathcal{N}_{\rm a}=64$ and $N_{\rm PD}=49$. A remarkable gap of $\sim18$ dB can be observed between MRC and EGC techniques (this is compatible with $10\log(N_{\rm PD})=18$ dB discussed in Section~\ref{SNRAnalysis}). This means that in order to attain the SNR performance of MRC by EGC, more number of outer arrays (i.e. a larger receiver) is required. However, this may not be applicable for mobile devices. In Fig.~\ref{Fig-SNR_MRC}, the analytical expression for the SNR of MRC given in \eqref{simplified-Av:SNR} is compared with the simulation results, which proves the accuracy of our analytical model. Hence one can rely on the proposed analytical model instead of OpticStudio ray-tracing simulations to save computation time.



\begin{figure}[t!]
	\centering  
	\subfloat[\label{sub1:VariableRateQAMEx1-optionII}]{%
		\resizebox{.24\linewidth}{!}{\includegraphics{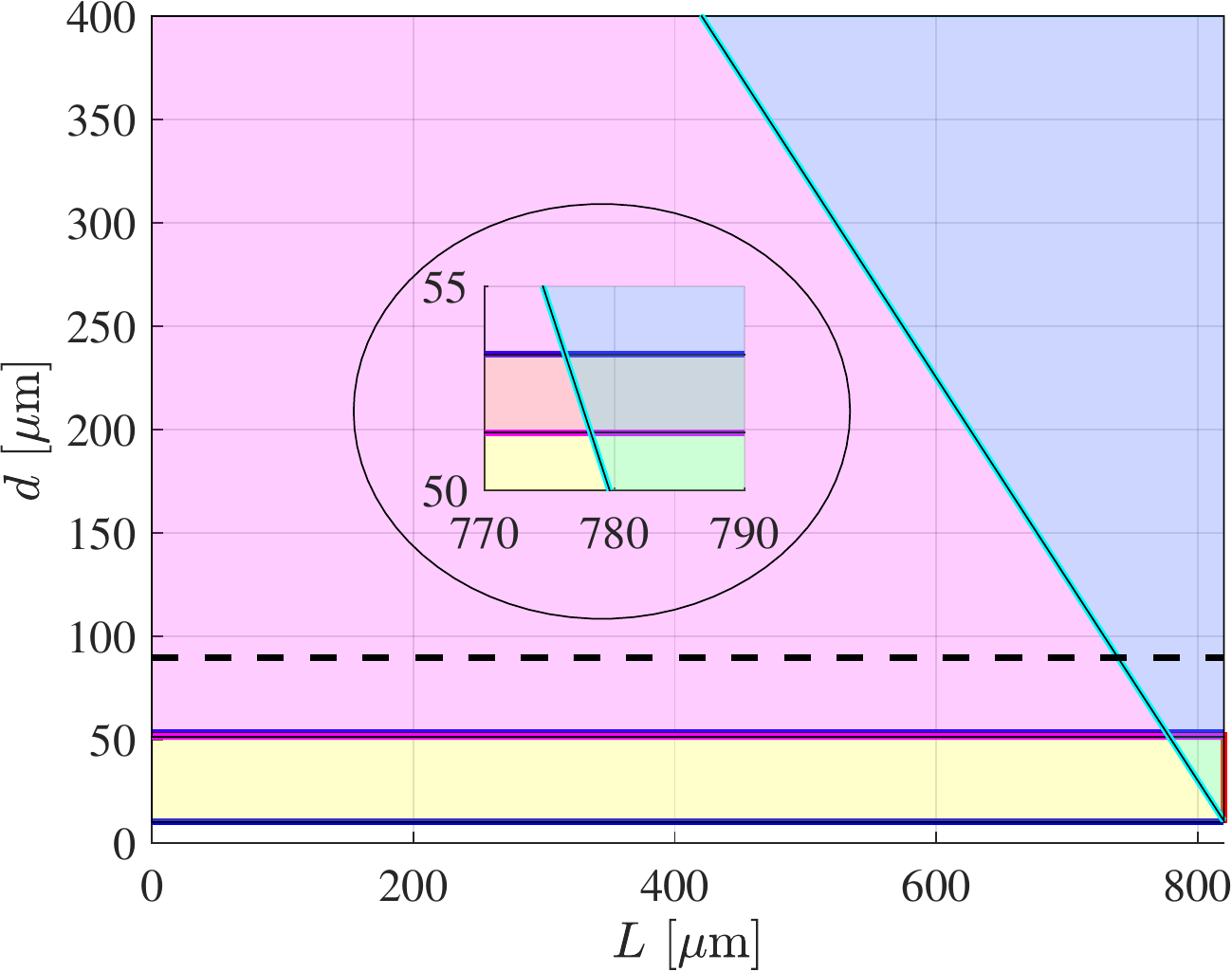}}
	}
	\subfloat[\label{sub2:VariableRateQAMEx1-optionII}]{%
		\resizebox{.24\linewidth}{!}{\includegraphics{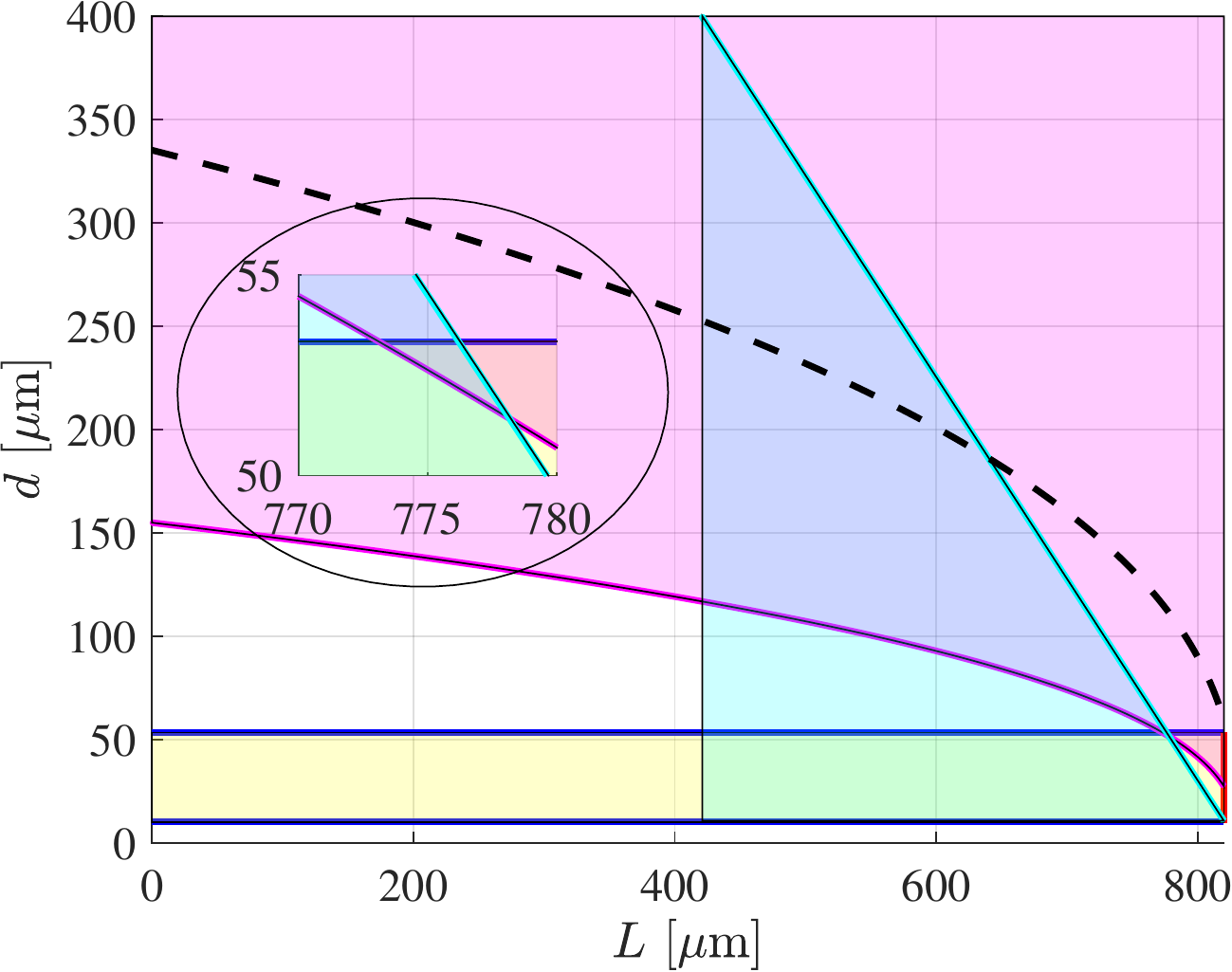}}
	}
	\subfloat[\label{sub3:VariableRateQAMEx1-optionII}]{%
		\resizebox{.24\linewidth}{!}{\includegraphics{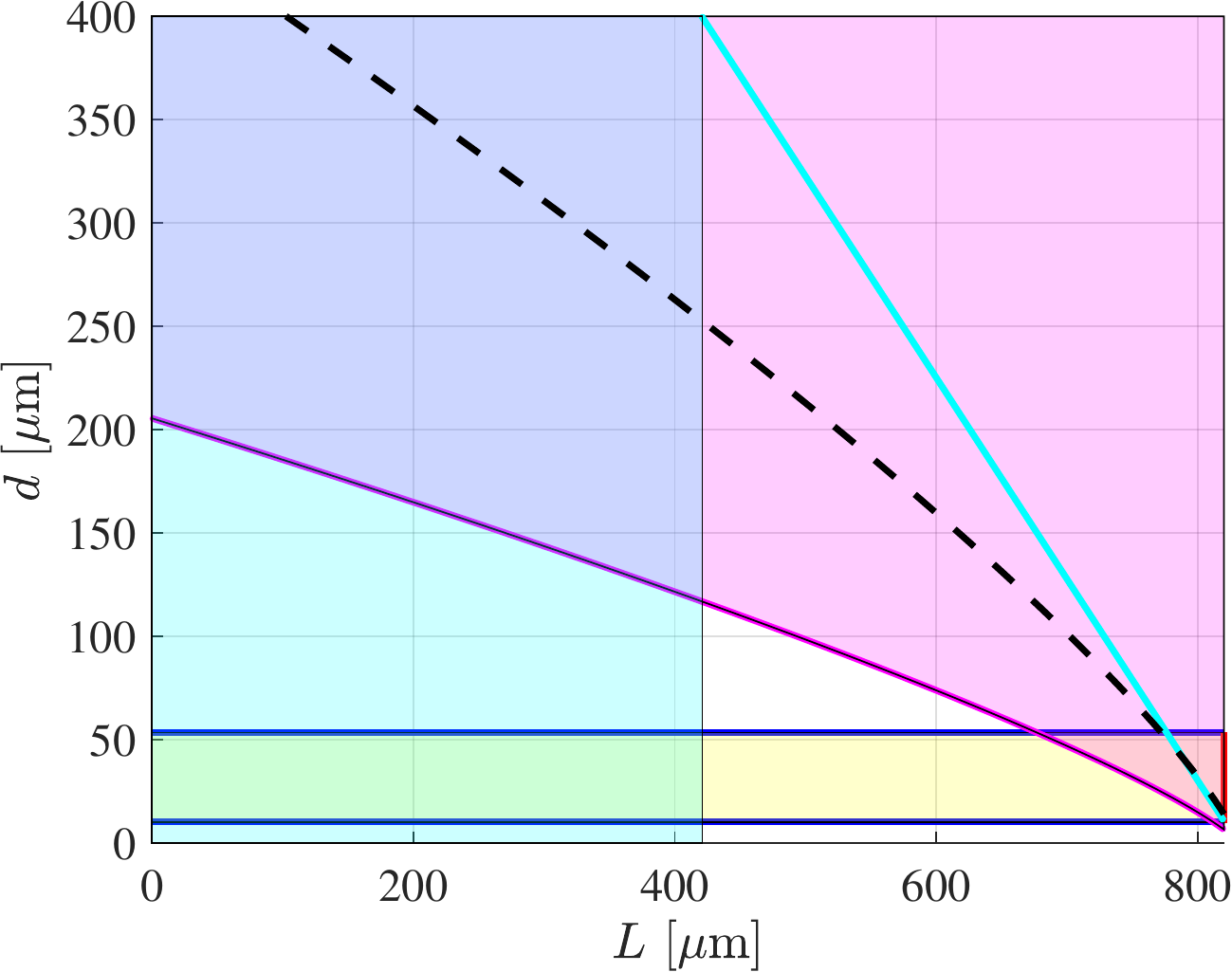}}
	}
		\subfloat[Data rate \label{sub3:VariableRateQAMEx1-OR-optionII}]{%
		\resizebox{.25\linewidth}{!}{\includegraphics{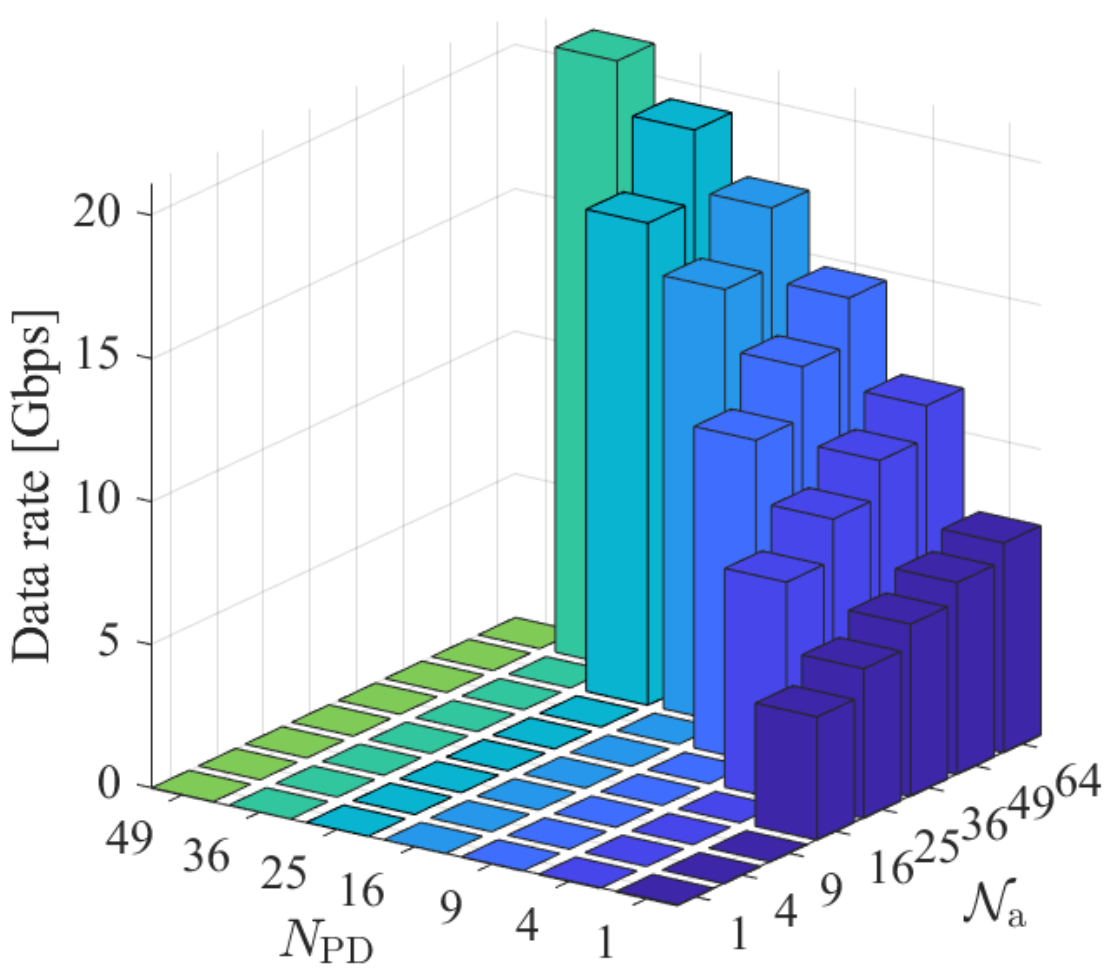}}
	}\\
	\caption{Feasible region based on (a) first, (b) second and (c) third equations of SNR given in \eqref{simplified-Av:SNR}.}
	\label{VariableRateQAMEx1-optionII}
	\vspace{-0.7cm}
\end{figure}

We now present the results of optimization problem for DCO-OFDM, where the number of subcarriers is set to $512$. 
Fig.~\ref{sub1:VariableRateQAMEx1-optionII}-c illustrate the feasible regions for $\gamma\geq \gamma_{\rm req}=10.6$ (this ensures ${\rm BER}\leq{\rm BER_{req}}=0.001$ for DCO-OFDM \cite{Kazemi2000PIMRC}) and ${\rm FOV}\geq{\rm FOV_{req}}=15^{\circ}$ (or equivalently for $L\leq 820\ \mu$m based on \eqref{Eq:L-FOV}), when $\mathcal{N}_{\rm a}=64$, $N_{\rm PD}=36$ and ${\rm{FF}}=0.64$. 
The cyan areas correspond to each sub-function of \eqref{simplified-Av:SNR}. The magenta areas represent the conditions given in \eqref{SNR-firstFunc}, \eqref{SNR-secondFunc} and \eqref{SNR-thirdFunc}. These conditions guarantee the required SNR. The yellow areas indicate the third constraint in \eqref{OP1}, which is $10\ \mu$m $\leq d \leq 400\sqrt{\frac{0.64}{{36}}}=53.33\ \mu$m.
The feasible region for each optimization problem can be found by looking into the intersection of different areas in each figure. It can be seen that there is no intersections between the areas in  Fig.~\ref{sub3:VariableRateQAMEx1-optionII}. However, the first equation of SNR in \eqref{simplified-Av:SNR} yields a nonempty feasible region as shown in Fig.~\ref{sub1:VariableRateQAMEx1-optionII}. There are plenty of combinations for the pair of $(d,L)$ that fulfil ${\rm BER}\leq0.001$ and ${\rm FOV}\geq15$. The optimum one that maximizes $R$ can be determined based on the third row of  Table~\ref{VariableRateSNR1-Table}. Since $R(d)$ is a monotonically increasing function in the range of $d\in[d_{\rm min},d_{\rm max}]$, $d_{\rm opt}\!=\!d_{\rm max}\!=\!53.33~\mu$m. Furthermore, $L_{\rm opt}\in [777,820]$ $\mu$m, where the choice of $L=777$ $\mu$m yields the \ac{FOV} of $16.4^{\circ}$. 
Fig.~\ref{sub3:VariableRateQAMEx1-OR-optionII} presents the achievable data rate versus $N_{\rm PD}$ in each inner array and $\mathcal{N}_{\rm a}$. 
A maximum data rate of $21.14$ Gbps can be achieved for the given BER of $0.001$ and the \ac{FOV} of $15^{\circ}$ using DCO-OFDM. It is also assumed that the maximum size of the receiver is $2$ cm $\times$ $2$ cm in these results. The inner array that provides the maximum data rate includes $6\times6$ PDs ($N_{\rm PD}=36$) with side length of $53.33~\mu$m. This size of a PD provides a bandwidth of $10$ GHz and it ensures the BER is less than $0.001$.
A minimum of $\mathcal{N}_{\rm a}=16$ arrays is required to guarantee the desired BER of $0.001$ and the required \ac{FOV} of $15^{\circ}$, which achieves data rate of $4.3$ Gbps.

\begin{figure}[t!]
	\centering  
	\subfloat[$\mathcal{N}_{\rm a} = 25$\label{sub5:RateVsLVRQAM}]{%
		\resizebox{.24\linewidth}{!}{\includegraphics{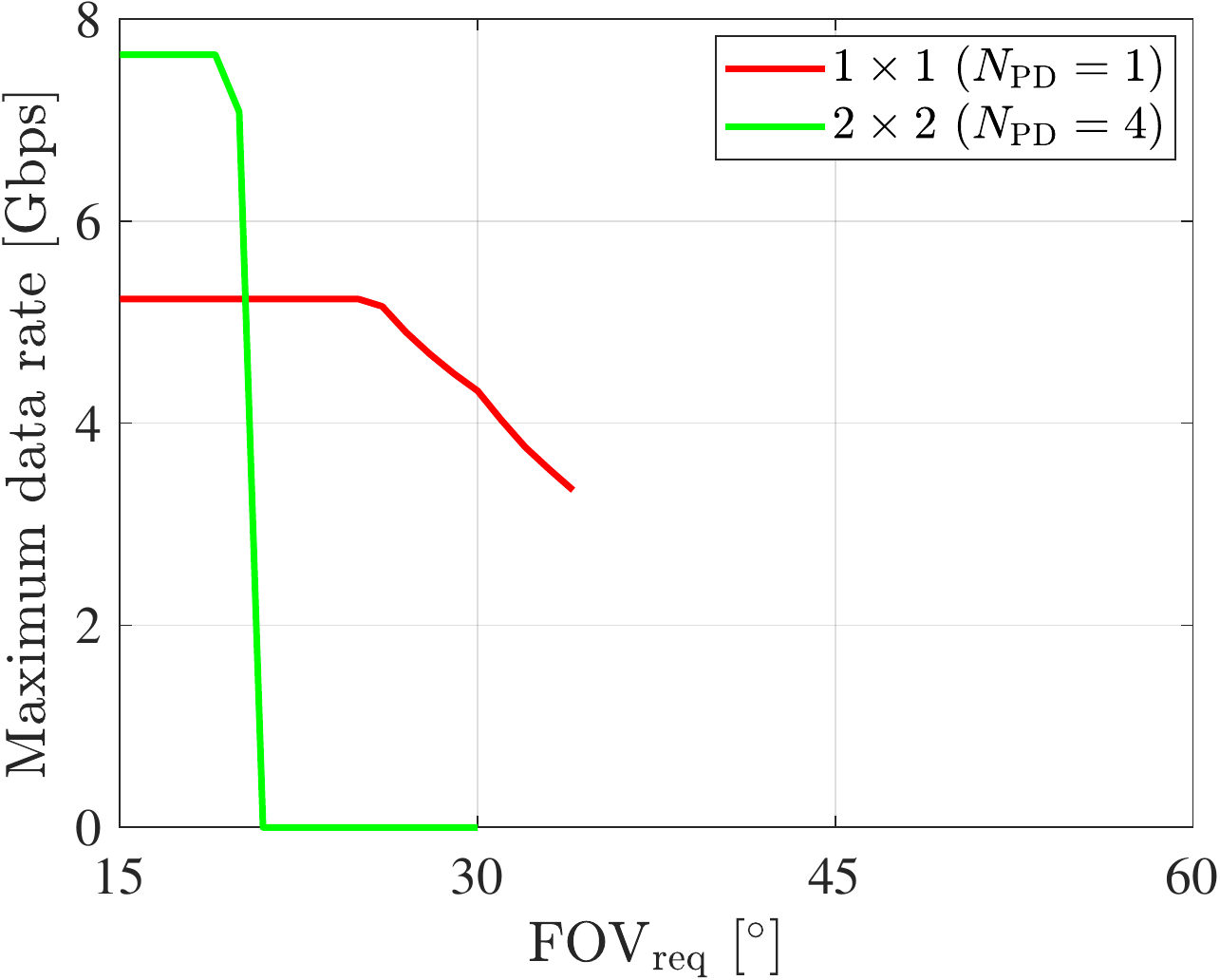}}
	}
	\subfloat[$\mathcal{N}_{\rm a} = 36$\label{sub6:RateVsLVRQAM}]{%
		\resizebox{.24\linewidth}{!}{\includegraphics{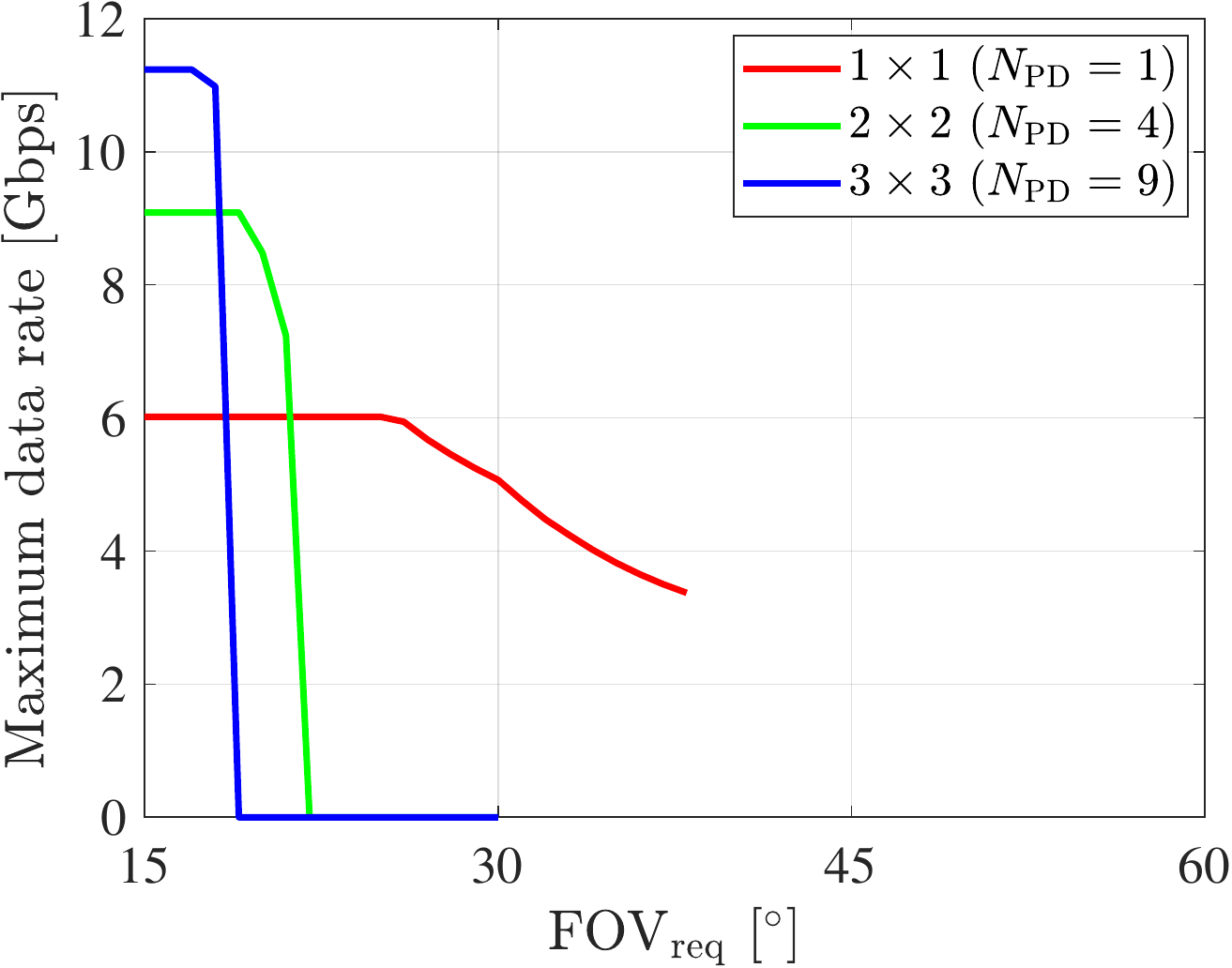}}
	}
	\subfloat[$\mathcal{N}_{\rm a} = 49$\label{sub7:RateVsLVRQAM}]{%
		\resizebox{.24\linewidth}{!}{\includegraphics{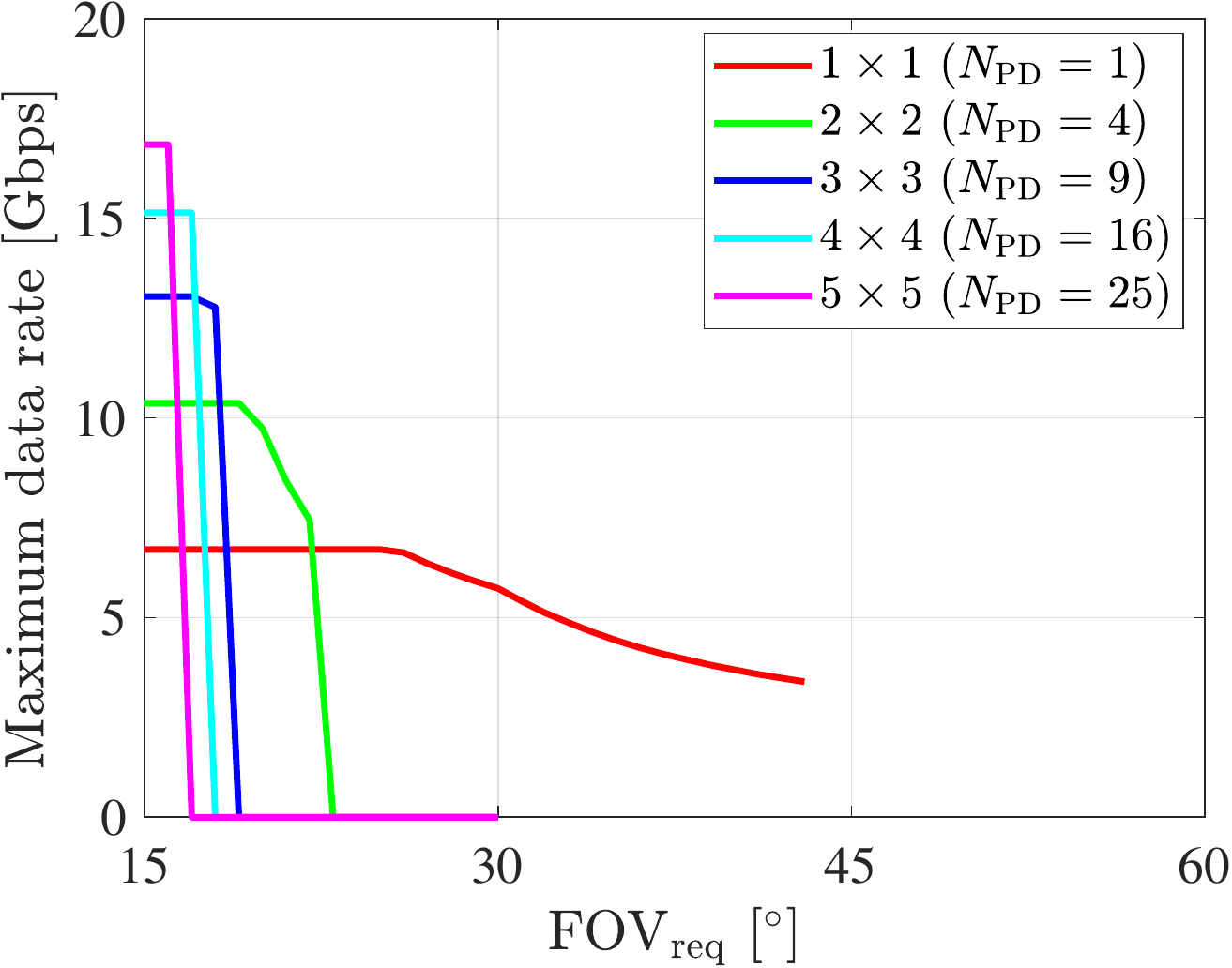}}
	}
	\subfloat[$\mathcal{N}_{\rm a} = 64$\label{sub8:RateVsLVRQAM}]{%
		\resizebox{.24\linewidth}{!}{\includegraphics{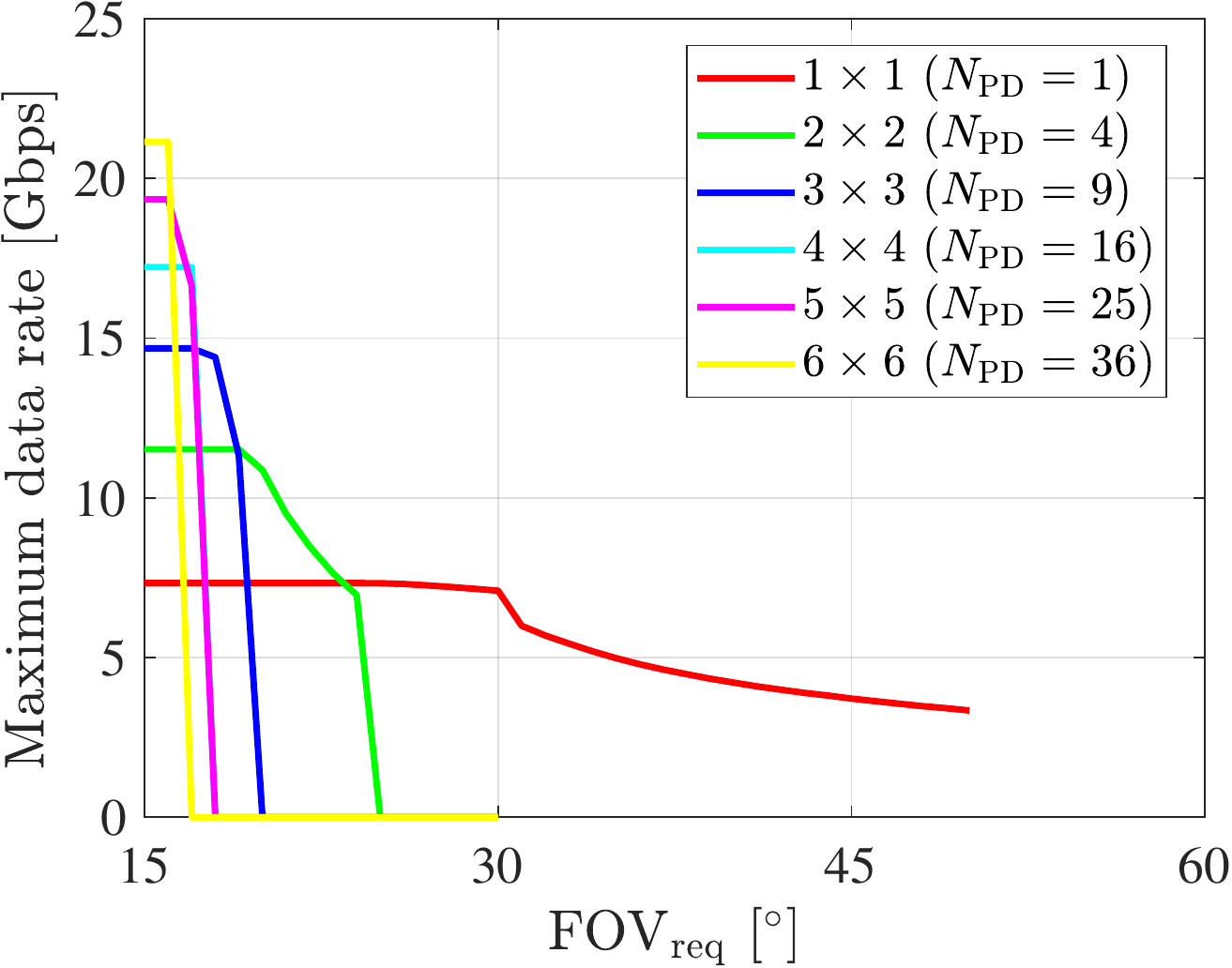}}
	}\\
	\caption{Maximum data rate versus $\rm{FOV_{req}}$ for different number of PDs on a single array and different sizes of array of arrays. DCO-OFDM and FF of $0.64$ are assumed for these results. }
	\label{RateVsLvariableRateQAM}
	\vspace{-0.7cm}
\end{figure}

Fig.~\ref{RateVsLvariableRateQAM} indicates the maximum data rate versus the required \ac{FOV} for DCO-OFDM. The rate-FOV trade-off can be observed in these results. Each subfigure represents a specific $\mathcal{N}_{\rm a}$. 
It should be noted that none of the receiver structures with $\mathcal{N}_{\rm a}=1$, $\mathcal{N}_{\rm a}=4$ and $\mathcal{N}_{\rm a}=9$ are able to fulfill the BER and \ac{FOV} requirements. A receiver with $\mathcal{N}_{\rm a}=16$ arrays can ensure the design requirements, only if each inner array includes $1$ PD.
As shown in Fig.~\ref{sub5:RateVsLVRQAM}, a receiver that includes $\mathcal{N}_{\rm a}=25$ arrays with either $N_{\rm PD}=1$ or $N_{\rm PD}=4$ structure for the inner arrays can ensure the desired BER and \ac{FOV}. The inner array with $N_{\rm PD}=4$ PDs support higher data rate and lower \ac{FOV} compared to the one with $N_{\rm PD}=1$ structure.
When the size of the receiver increases to  $\mathcal{N}_{\rm a}=64$ arrays, each inner array with $1$ PD upto $36$ PD arrangement are able to support BER of less than $0.001$ and the required \ac{FOV}. The maximum data rate is related to $\mathcal{N}_{\rm a}=64$ arrays with $N_{\rm PD}=36$ PDs on each inner array, which is $21.14$ Gbps. These maximum data rate can be achieved for $777 \leq L \leq 820$ $\mu$m.  

\begin{figure}[t!]
	\centering  
	\subfloat[\label{sub1:OOKOptimizationEx1}]{%
		\resizebox{.24\linewidth}{!}{\includegraphics{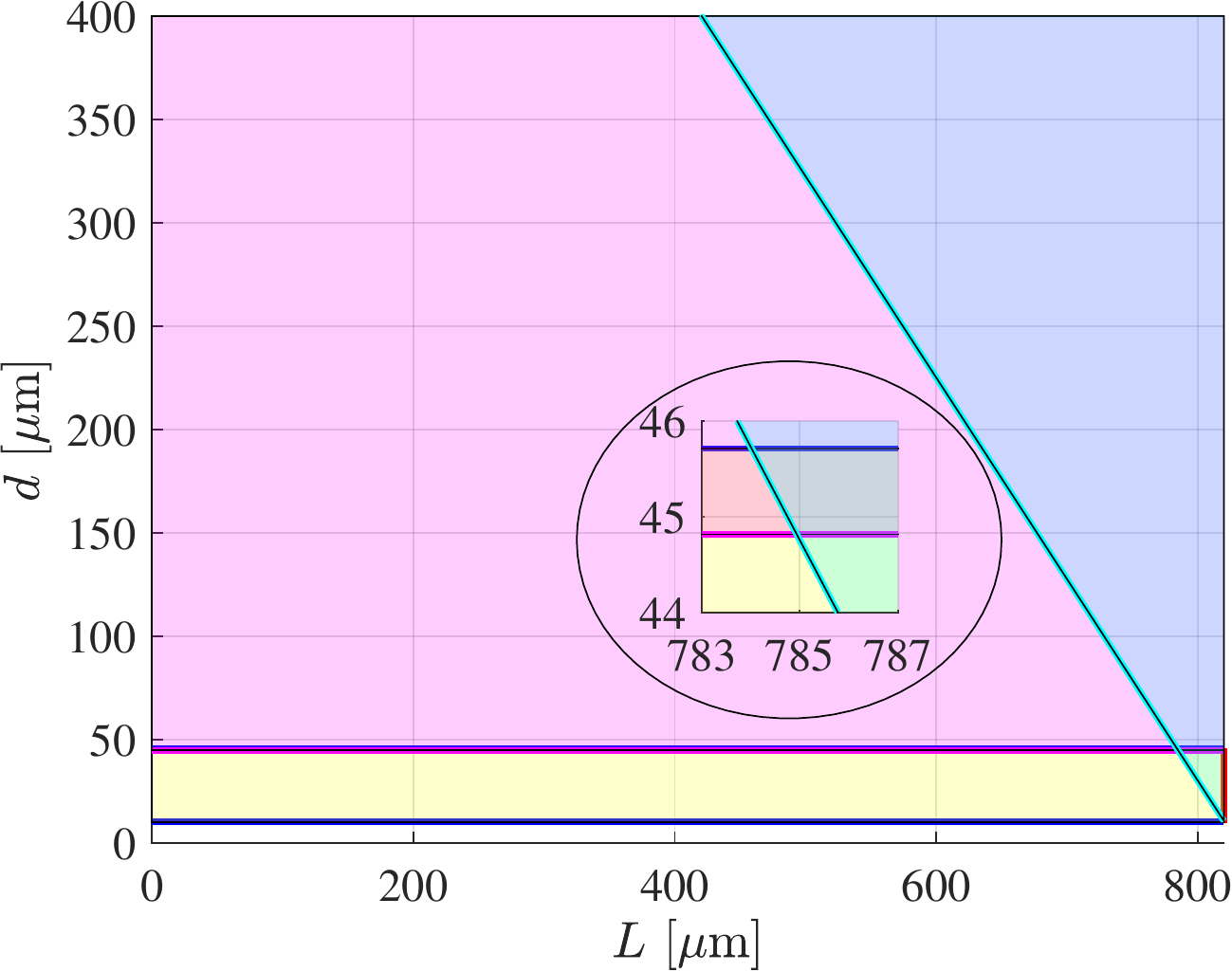}}}\
	\subfloat[\label{sub2:OOKOptimizationEx1}]{%
		\resizebox{.24\linewidth}{!}{\includegraphics{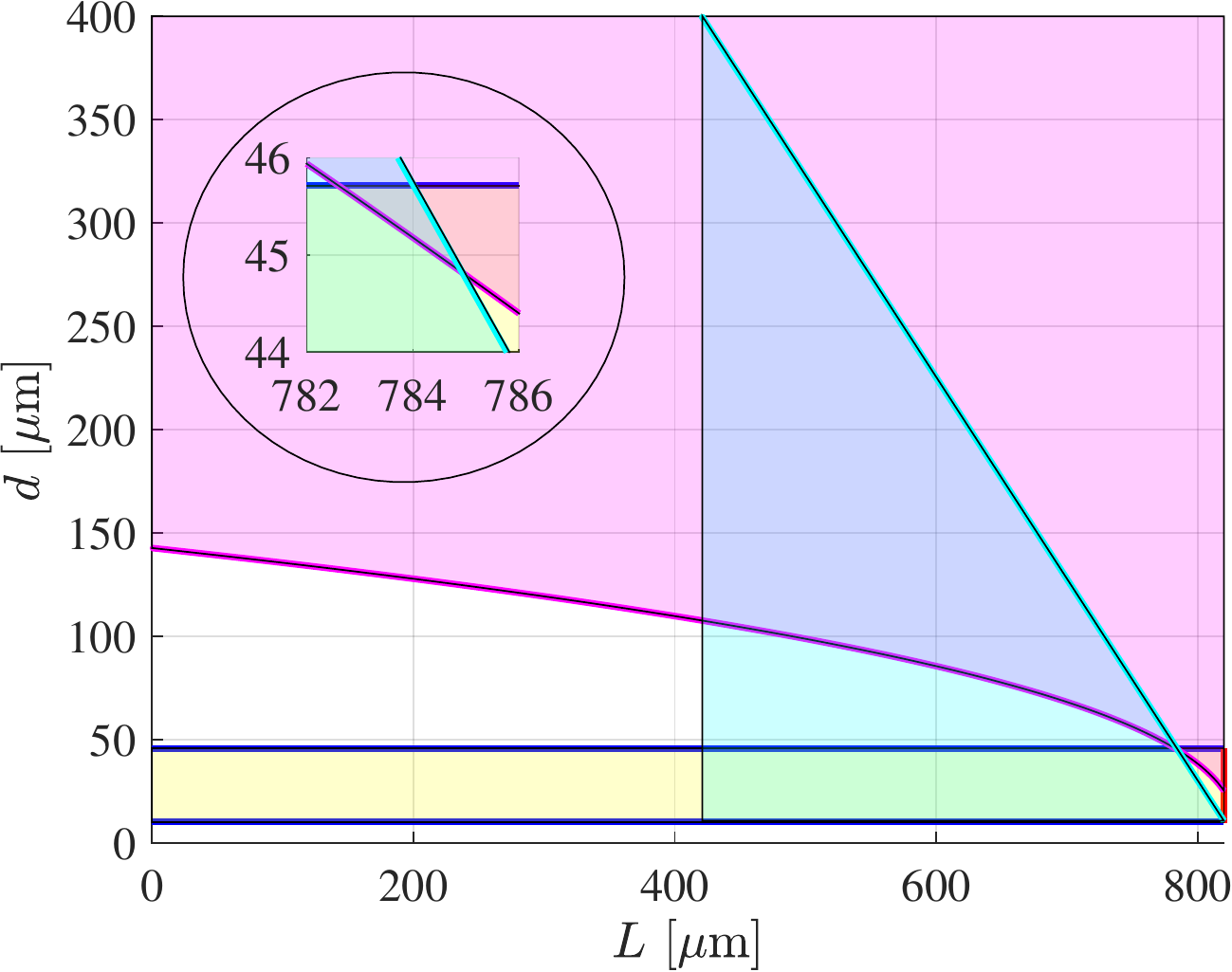}}}\
	\subfloat[\label{sub3:OOKOptimizationEx1}]{%
		\resizebox{.24\linewidth}{!}{\includegraphics{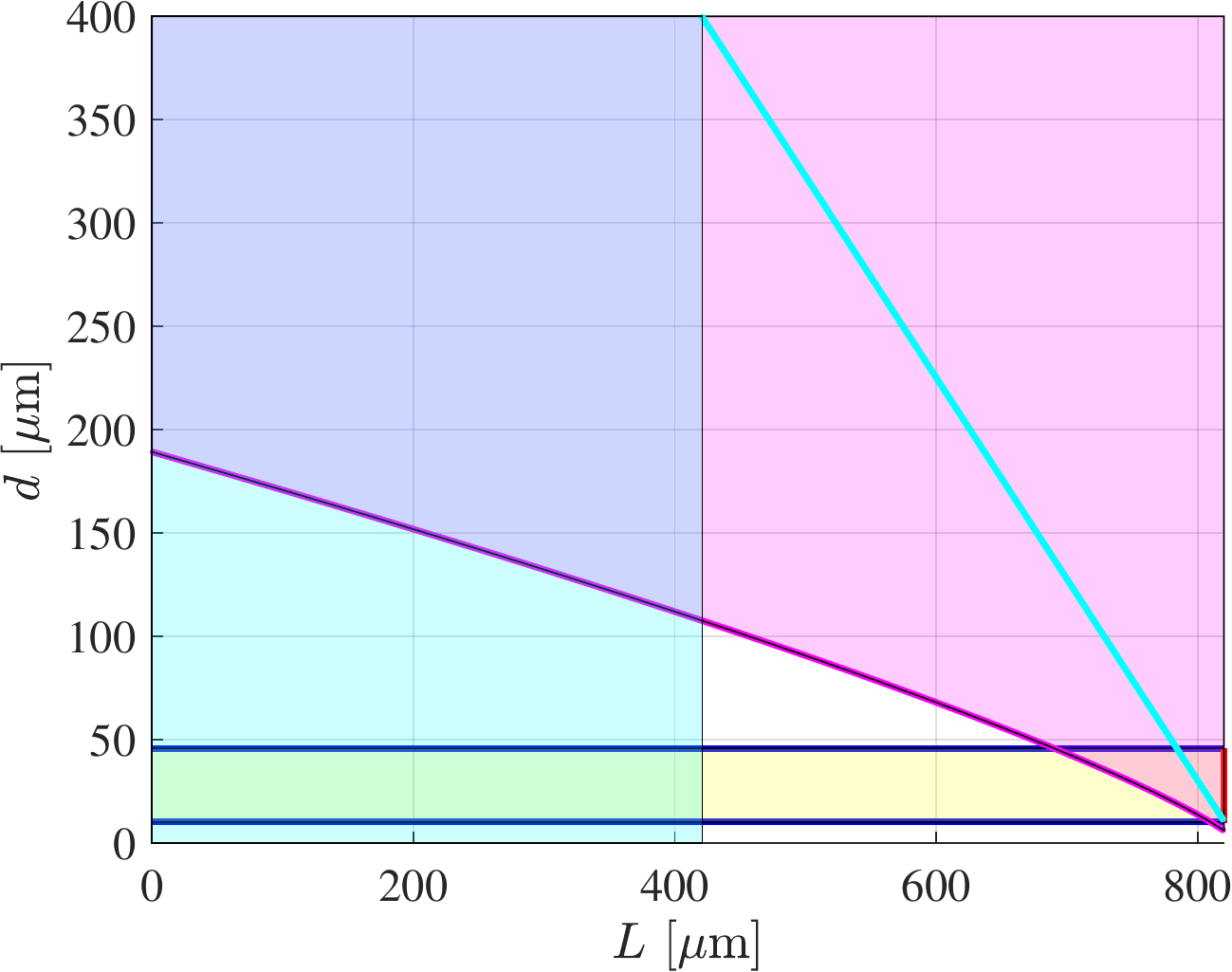}}}\
	\subfloat[Data rate \label{sub3:OOKOptimizationEx1-OR}]{%
		\resizebox{.25\linewidth}{!}{\includegraphics{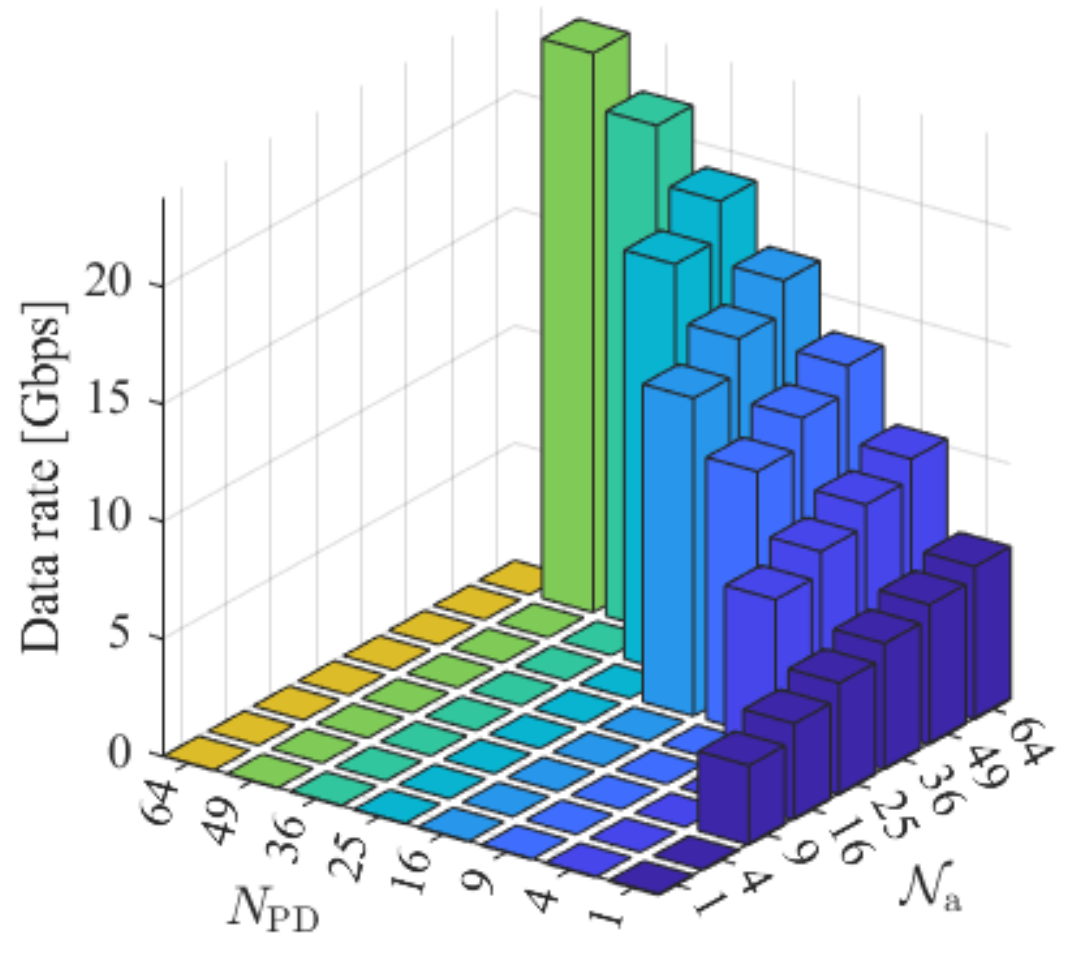}}}\\
	\caption{Feasible region based on (a) first, (b) second and (c) third equations of SNR given in \eqref{simplified-Av:SNR}.}
	\label{OOKOptimizationEx1}
	\vspace{-0.7cm}
\end{figure}

Next, we present the results for OOK. 
Fig.~\ref{OOKOptimizationEx1} illustrates these feasible regions for optimization problems given in \eqref{OP1-General}, \eqref{OP2-General} and \eqref{OP3-General} for $\gamma\geq \gamma_{\rm req}=9.55$ and ${\rm FOV}\geq{\rm FOV_{req}}=15^{\circ}$, when 
$\mathcal{N}_{\rm a}=64$, $N_{\rm PD}=49$ and ${\rm{FF}}=0.64$. The constraint $\gamma\geq \gamma_{\rm req}=9.55$ ensures ${\rm BER}\leq{\rm BER_{req}}=10^{-3}$ for the OOK modulation \cite{Kahn97WirelessInfrared}. 
%
The yellow areas indicate the third constraint in \eqref{OP1}, which is $10\ \mu$m $\leq d \leq 400\sqrt{\frac{0.64}{{49}}}=45.7\ \mu$m. The feasible region for the optimization problem can be found by looking into the intersection of the areas. The subsets in Fig.~\ref{sub1:OOKOptimizationEx1} and Fig.~\ref{sub2:OOKOptimizationEx1} depict the zoomed intersection areas for convenience. The optimum value of $d$ is the smallest in the intersection area to ensure the maximum data rate since $R=\frac{2}{C_{\rm t}d}$. Therefore, $d_{\rm opt}=d_{\Delta}=44.81$~$\mu$m and $L_{\rm opt}\in[785,820]$ $\mu$m, which are in agreement with the analytical solutions given in first row of the Table~\ref{OOK-OP-Solution} for OOK modulation. We note that the choice of $L=785~\mu$m results in the \ac{FOV} of $16.2^{\circ}$.  
Fig.~\ref{sub3:OOKOptimizationEx1-OR} presents the achievable data rate versus the number of PDs in each inner array and the number of arrays. 
A maximum data rate of $23.82$ Gbps can be achieved for a given BER of $10^{-3}$ and a \ac{FOV} of $15^{\circ}$ using OOK modulation. It is assumed that the maximum length of the receiver is $2$ cm in these results. The inner array that provides the maximum data rate includes $7\times7$ PDs each with a side length of $44.81\ \mu$m. This size of a PD provides a bandwidth of $11.91$ GHz and it ensures the BER is less than $10^{-3}$. The achieved FF of the array is $0.61$ which is very close to our target FF of $0.64$.
A minimum of $\mathcal{N}_{\rm a}=9$ is required to fulfill the required BER of $10^{-3}$ and the required \ac{FOV} of $15^{\circ}$, which achieves data rate of $3.38$ Gbps. It can be also observed that when the number of PDs ($N_{\rm PD}$) increases, a bigger size of receiver (i.e., larger $\mathcal{N}_{\rm a}$) is also required to satisfy the BER and FOV requirements. This is due the higher thermal noise accumulated in a larger PD array.

\begin{figure}[t!]
	\centering 
	\subfloat[$\mathcal{N}_{\rm a} = 25$ \label{sub5:RateVsOOKL}]{%
		\resizebox{.23\linewidth}{!}{\includegraphics{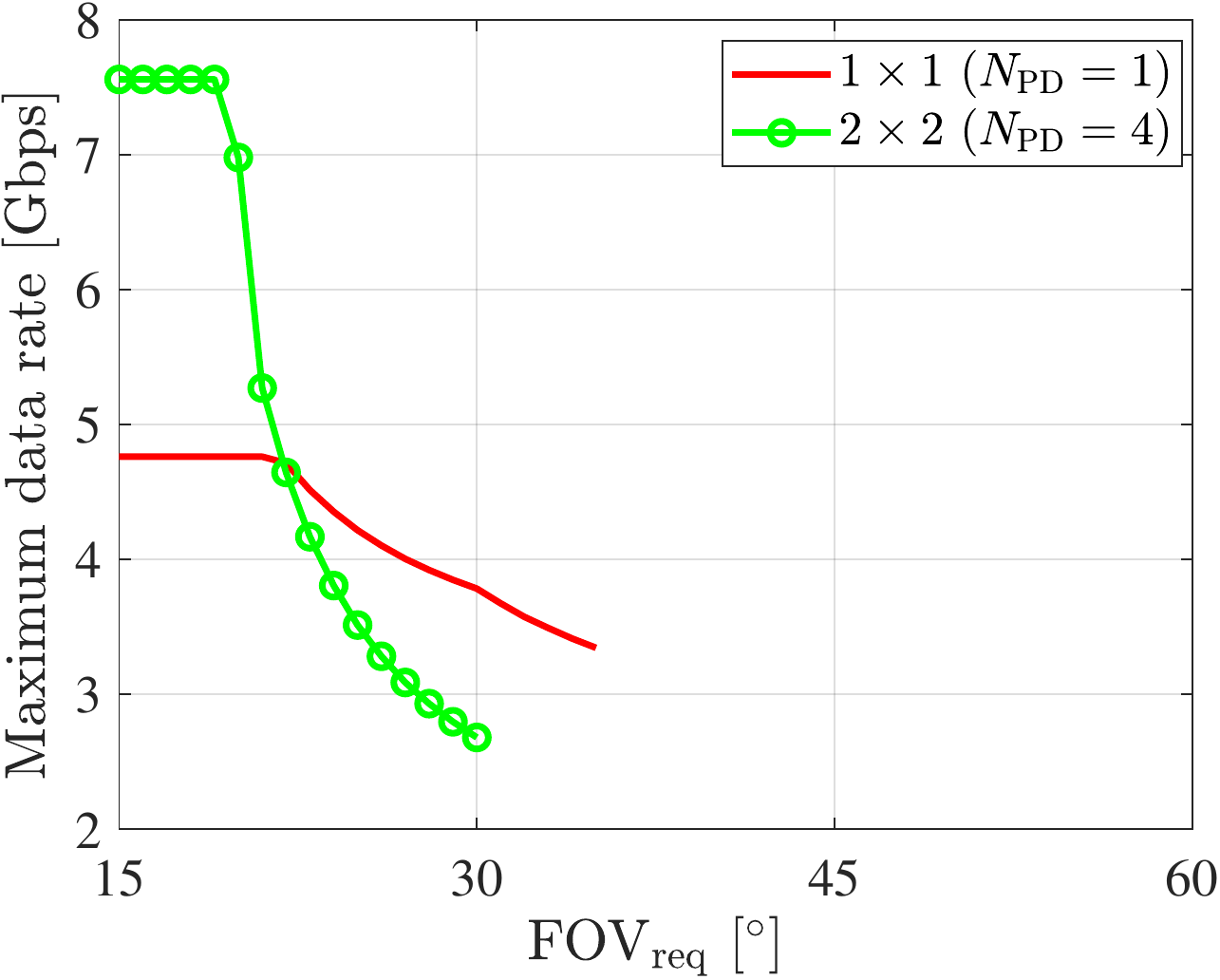}}}\ \
	\subfloat[$\mathcal{N}_{\rm a} = 36$ \label{sub6:RateVsLOOK}]{%
		\resizebox{.23\linewidth}{!}{\includegraphics{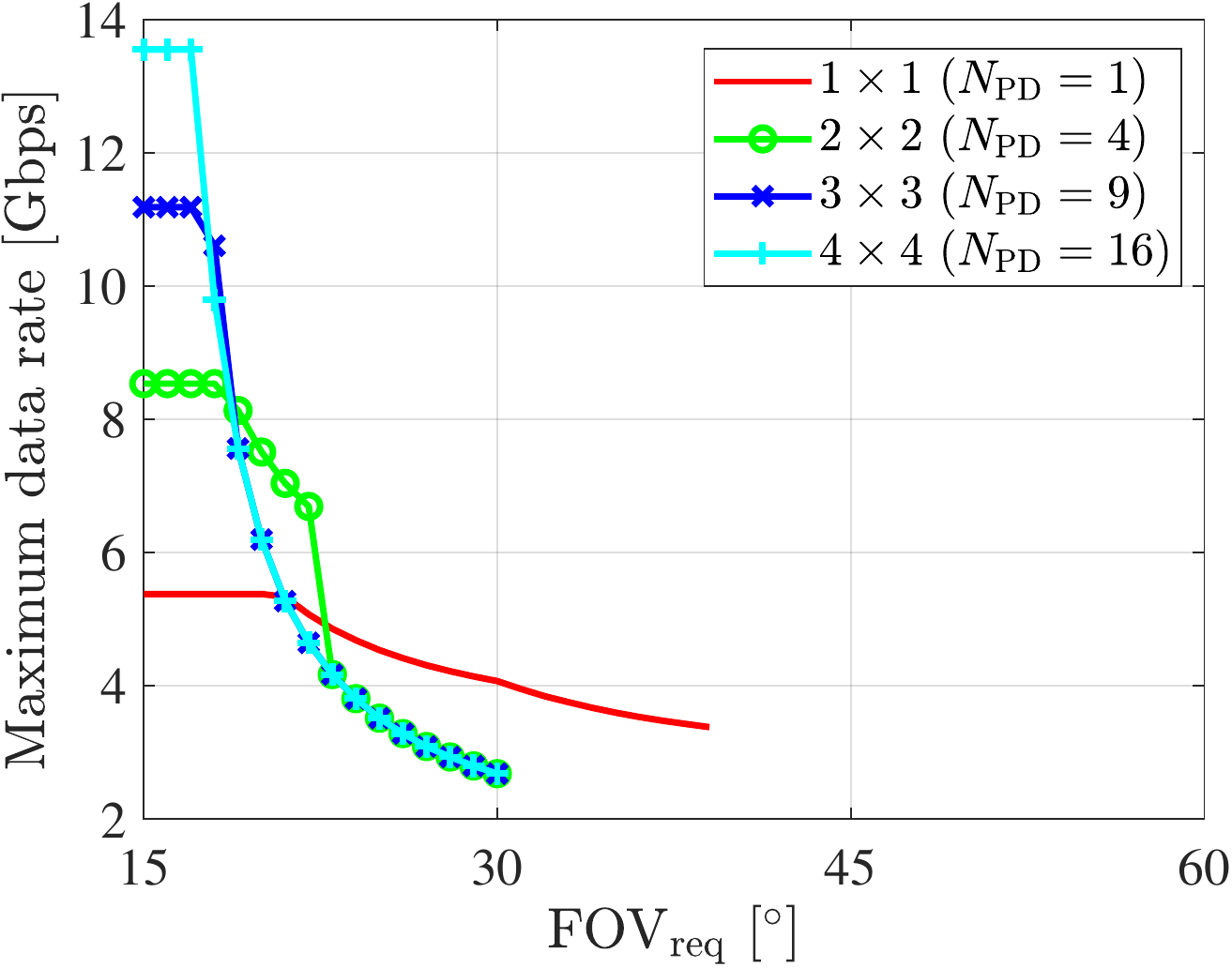}}}\ \
	\subfloat[$\mathcal{N}_{\rm a} = 49$ \label{sub7:RateVsLOOK}]{%
		\resizebox{.23\linewidth}{!}{\includegraphics{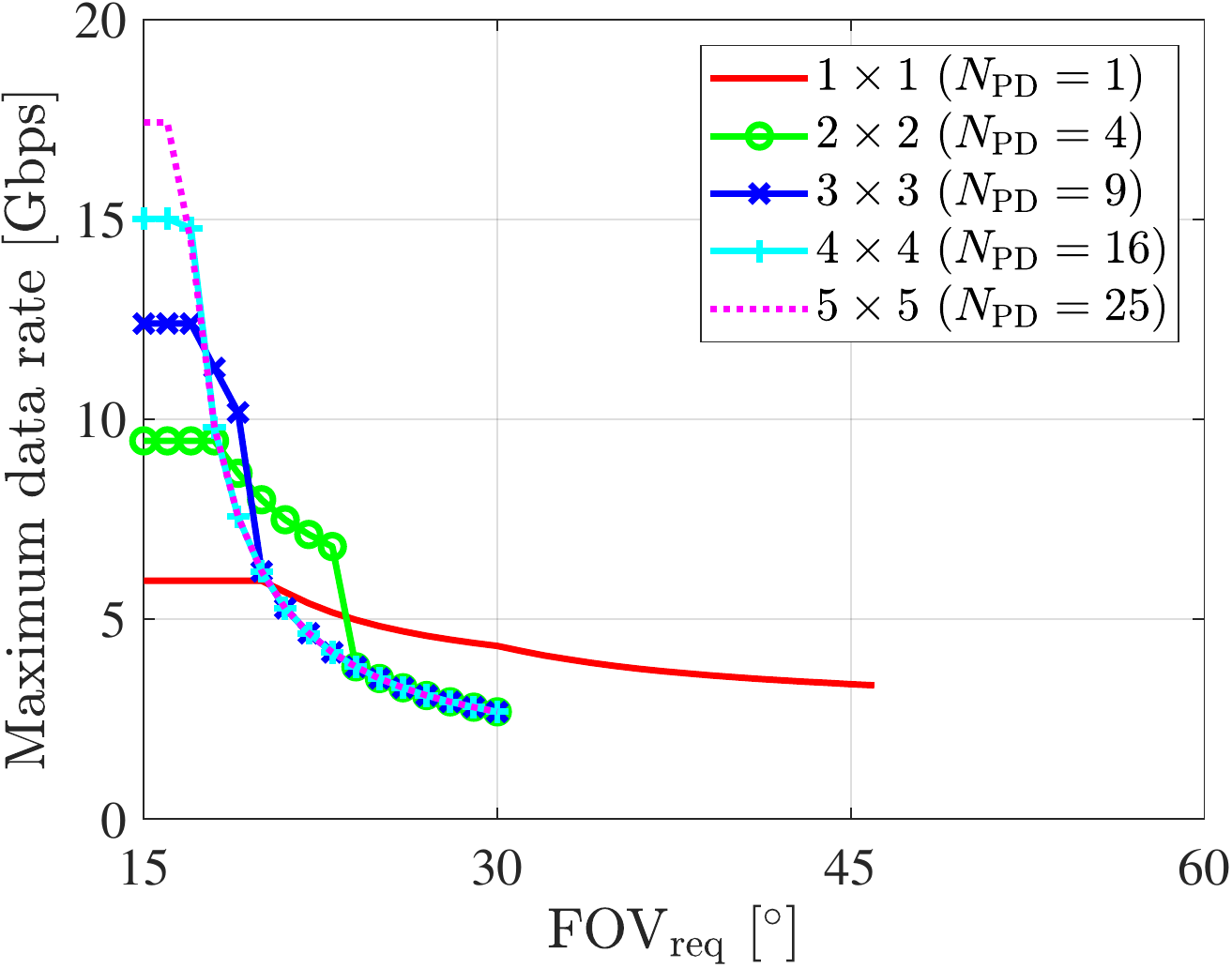}}}\ \
	\subfloat[$\mathcal{N}_{\rm a} = 64$ \label{sub8:RateVsLOOK}]{%
		\resizebox{.23\linewidth}{!}{\includegraphics{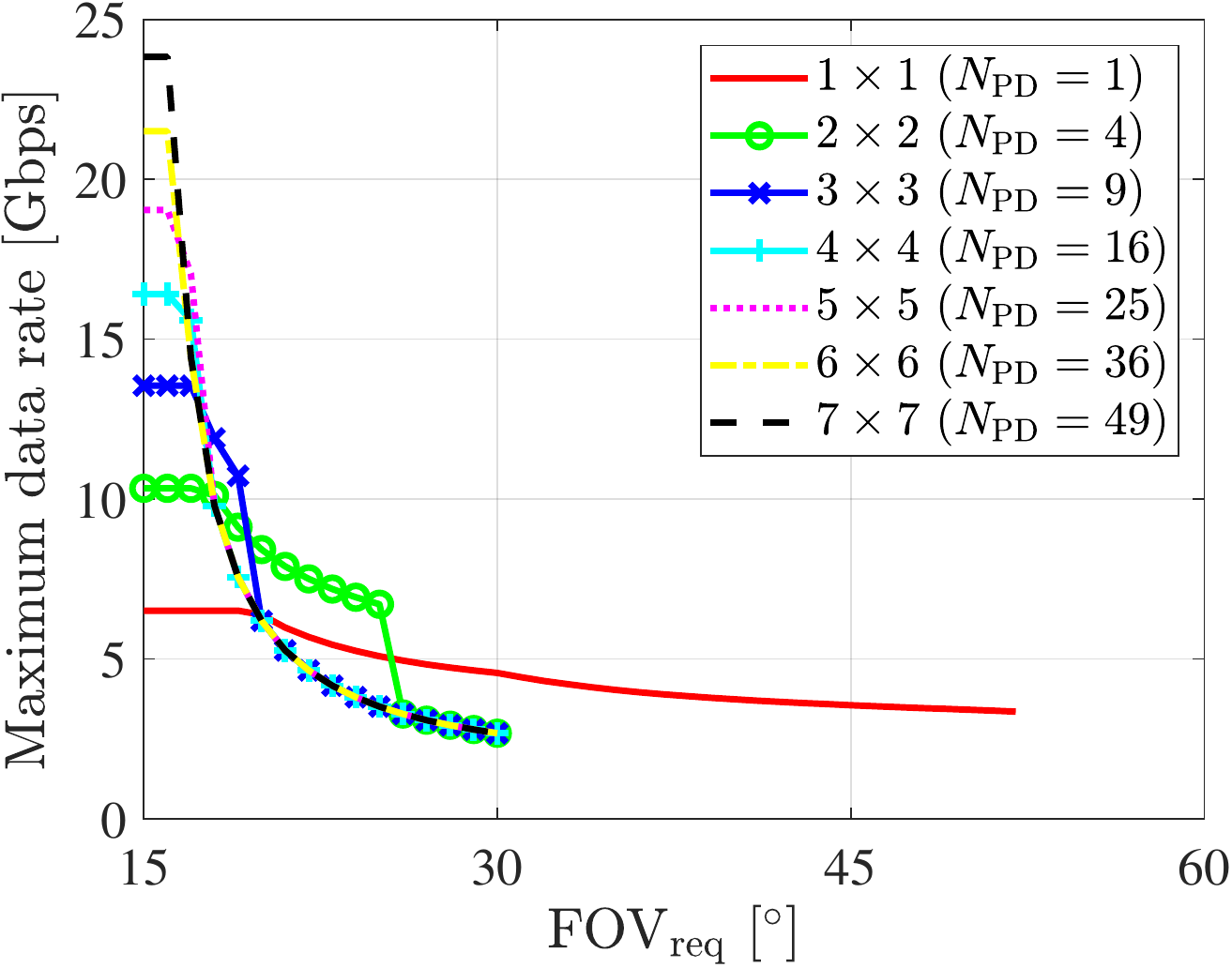}}}\\
	\caption{Maximum data rate versus $\rm{FOV_{req}}$ for different number of PDs on the inner array and different values of $\mathcal{N}_{\rm a}$. OOK modulation and FF of $0.64$ are assumed for these results.}
	\label{RateVsLOOK}
	\vspace{-0.7cm}
\end{figure}

Fig.~\ref{RateVsLOOK} show the maximum data rate versus the required \ac{FOV}. 
Each subfigure represents a specific size of receiver (indicated by the number of outer arrays, $\mathcal{N}_{\rm a}$). 
We note that neither $\mathcal{N}_{\rm a}=1$ nor $\mathcal{N}_{\rm a}=4$ are able to fulfill the BER and \ac{FOV} requirement. Both sizes of $\mathcal{N}_{\rm a}=9$ and $\mathcal{N}_{\rm a}=16$ can ensure the design requirements, only if $N_{\rm PD}=1$.
A receiver with $\mathcal{N}_{\rm a}=25$ with either $N_{\rm PD}=1$ or $N_{\rm PD}=4$ configuration is able to guarantee BER of less than $0.001$ and required \ac{FOV} of $15^{\circ}$ as shown in Fig.~\ref{sub5:RateVsOOKL}. 
The rate-\ac{FOV} trade-off can be observed in these results, where higher data rate are achievable in cost of lower \ac{FOV}. 
For $\mathcal{N}_{\rm a}=25$, the PD configuration with $N_{\rm PD}=4$ is able to fulfill $\rm{FOV_{req}}\leq 30^{\circ}$ while the PD configuration with $N_{\rm PD}=1$ can support $\rm{FOV_{req}}\leq35^{\circ}$. 
As shown in Fig.~\ref{sub8:RateVsLOOK}, when $\mathcal{N}_{\rm a}$ increases to $64$, inner arrays with $N_{\rm PD}$ up to $49$ detectors are able to support the desired BER and \ac{FOV}. The maximum data rate is related to the receiver structure with $\mathcal{N}_{\rm a}=64$ and $N_{\rm PD}=49$, which is $23.82$ Gbps. These maximum data rate can be achieved for $785 \leq L \leq 820$ $\mu$m. 
It worth mentioning that inner arrays with $N_{\rm PD}\geq 64$ are not able to satisfy the BER requirement. The reason is that the high bandwidth of the PDs adds more noise to the system. This is in accordance with the early statement that choosing high bandwidth PDs in a power-limited regime is not helpful as more noise will be added to the system.
It is also noted that $\rm{FOV}\geq 30^{\circ}$ can be achieved only with single-PD arrays at very close distances to the lens. Arrays with other PD configurations fail to assure $\rm{BER}\leq0.001$. 
For a single PD equal to the size of the array  ($400\times400$ $\mu$m$^2$), we can attain $70^{\circ}$ \ac{FOV} at $L\approx0$ (see Fig.~\ref{Fig:L-FoV}) and a maximum data rate of $2.68$ Gbps with $\mathcal{N}_{\rm a}=64$. 
It should be highlighted that we are in a power-limited regime due to eye safety considerations, where lower modulations order are preferable.
That is the reason, OOK outperforms DCO-OFDM. However, in a high SNR regime (which can be realized with APD), DCO-OFDM should be able to outperform OOK. 

\vspace{-0.1cm}
\section{Summary and Concluding Remarks}
\label{Sec:Summary}
\vspace{-0.1cm}
In this study, we proposed an imaging receiver which consists of an array of arrays structure to deliberately address the two well-known trade-offs between area-bandwidth and gain-FOV. The proposed receiver is able to control the combined impacts of both trade-offs effectively achieving different required performances defined by a rate-FOV trade-off. We derived the analytical models of SNR for both EGC and MRC techniques for the proposed array of arrays structure. The analytical models are verified with realistic simulations conducted in OpticStudio (Zemax) software.
We then formulated an optimization problem to assure design requirements such as \ac{FOV} and BER for OOK modulation and DCO-OFDM. Furthermore, we derived the optimum side length of PDs and the distance between an array and the lens to maximize the achievable data rate of the system. In order to have realistic simulations, we considered practical aspects of receiver elements, such as the transmission coefficient of the lens, and so on. The simulations of beam intensity profile and SNR are carried out using the OpticStudio software, which ensures the reliability of our results. Insightful results and in-depth discussions are demonstrated for both OOK and DCO-OFDM. 
Simulation results confirm that with a square lattice arrangement of $8\times 8$ array of arrays, we are able to achieve a maximum data rate of $23.82$~Gbps and $21.14$~Gbps data rate with a FOV of $15^{\circ}$ using OOK and DCO-OFDM, respectively. Some possible directions for future research can be the use of a non-imaging structure, angle diversity receiver formation and selection combining technique. 

\vspace{-0.2cm}
\section{Acknowledgement}
The authors acknowledge financial support from the EPSRC under program grant EP/S016570/1 ``Terabit Bidirectional Multi-User Optical Wireless System (TOWS) for 6G LiFi''.

\vspace{-0.3cm}
\bibliography{reference}

\end{document}